\documentclass{article}
\usepackage{jheppub}
\usepackage[utf8]{inputenc}
\usepackage[T1]{fontenc}
\usepackage{graphicx} 
\usepackage{amsmath}
\usepackage{amstext}
\usepackage{amssymb,amsthm}
\usepackage{mathrsfs}
\usepackage{braket}
\usepackage{enumerate}
\usepackage{color}
\usepackage[numbers,sort&compress]{natbib}
\usepackage[normalem]{ulem}

\numberwithin{equation}{section}
\newcommand{\HH}{\mathcal{H}}
\newcommand{\U}{\mathcal{U}}
\newcommand{\A}{\mathcal{A}}
\newcommand{\B}{\mathcal{B}}

\makeatletter

\newcommand{\Rmnum}[1]{\rm \expandafter\@slowromancap\romannumeral #1@}
\makeatother

\newcommand{\ev}[2]{\bra{#1}#2\ket{#1}}

\newcommand{\M}{\mathcal{M}}
\newcommand{\Z}{\mathcal{Z}}
\renewcommand{\c}{\mathcal{C}}
\renewcommand{\d}{\mathcal{D}}
\newcommand{\R}{\mathbb{R}}
\renewcommand{\r}{\mathcal{R}}
\newcommand{\C}{\mathbb{C}}

\renewcommand{\O}{\mathcal{O}}
\newcommand{\hn}{\hat{n}}
\newcommand{\inn}[2]{\langle #1|#2\rangle}
\renewcommand{\tilde}[1]{\widetilde{#1}}

\title{An algebra for covariant observers in de Sitter space}
\author[1,2,3]{Bin Chen}
\author[2]{Jie Xu}
\affiliation[1]{Institute of Fundamental Physics and Quantum Technology, \\
\&School of Physical Science and Technology, Ningbo University, Ningbo,
 Zhejiang 315211, China}
\affiliation[2]{School of Physics, Peking University, No.5 Yiheyuan Rd, Beijing 100871, P. R. China}
\affiliation[3]{Center for High Energy Physics, Peking University, No.5 Yiheyuan Rd, Beijing 100871,
 P. R. China}
 \emailAdd{chenbin1@nbu.edu.cn, jiexu@stu.pku.edu.cn}

\abstract{
In $d$-dimensional de Sitter spacetime, consistency of the perturbative expansion necessitates imposing all second-order gravitational constraints associated with the $SO(1,d)$ isometry group, rather than restricting to the $\R\times SO(d-1)$ subgroup, to address linearization instability \cite{DeVuyst:2024grw}. Since generic de Sitter isometries do not preserve a fixed static patch, these constraints cannot be implemented within a fixed local algebra.
In this paper, we develop a framework that consistently imposes all $SO(1,d)$ constraints while incorporating multiple observers on arbitrary timelike geodesics. This is achieved by introducing the concept of \textit{covariant observer},  whose geodesic transforms covariantly under the isometry group. Upon quantization, the observer is described by a superposition of geodesics, with the associated static patches fluctuating, providing a quantum reference frame $L^2(SO(1,d))$. We realize this structure in an action model in which a particle carries a set of conserved charges, each one corresponding to a generator of de Sitter isometry group, which parametrize its geodesic and upon quantization lead to a fluctuating geodesic.

Inspired by the timelike tube theorem, we propose that the algebra of observables accessible to a covariant observer is generated by all degrees of freedom within its fluctuating static patch, including quantum field modes and other observers, which are treated as part of the matter system. Imposing the $SO(1,d)$ constraints yields a gauge-invariant algebra that takes the form of an averaged modular crossed product algebra over static patches and configurations of other geodesics, thereby generalizing the notion of a local algebra associated with a fixed region to that of a \textit{fluctuating region}. We show this algebra is of type \Rmnum{2} by explicitly constructing a faithful normal trace, leading to an observer-dependent notion of von Neumann entropy. For semiclassical states, by imposing a UV cutoff in QFT and proposing a quantum generalization of the first law, we demonstrate the agreement between the algebraic and generalized entropies. }

\begin{document}
\maketitle
\flushbottom

\section{Introduction}
As classical general relativity is a generally covariant theory with diffeomorphisms as  gauge symmetries, any physical observable should be relationally defined, that is, one chooses a dynamical reference frame and then describes  all other degrees of freedom relative to this frame in a gauge-invariant fashion \cite{Goeller:2022rsx}. Treating  diffeomorphism as gauge symmetry also presents a fundamental challenge  for quantum gravity (QG), and has led to several important insights. In particular, imposing the second-order constraints associated with the boost symmetry within the linearized theory has been shown to intrinsically regularize entanglement entropy, without the need to introduce an explicit ultraviolet (UV) regulator \cite{Witten:2021unn,Chandrasekaran:2022eqq,Chandrasekaran:2022cip}. \footnote{For an excellent review, see \cite{Liu:2025krl}.}

The ultraviolet divergence of entanglement entropy in quantum field theory arises from infinitely many entangled modes across the boundary of a local region.  This divergence is reflected in the  universal type \Rmnum{3}${}_1$ nature of the associated observable algebras \cite{Fredenhagen:1984dc,Buchholz:1986bg,Buchholz:1995gr,Haag:1996hvx,Yngvason:2004uh,Witten:2018zxz}, which lack a well-defined trace and therefore preclude an intrinsic definition of algebraic entropy \cite{takesaki1979theory,Sorce:2023fdx}.  This algebraic structure persists in semiclassical quantum gravity,  namely quantum field theory in curved spacetime, and underlies the difficulty of defining entropy in gravitational settings.

A concrete manifestation of this \Rmnum{3}${}_1$ structure arises in AdS/CFT holography \cite{Leutheusser:2021qhd,Leutheusser:2021frk}. The \Rmnum{3}${}_1$ algebra emerges from the large $N$ limit of the boundary algebra of single-trace operators in one copy of a CFT prepared in the thermofield double state at a temperature above the Hawking-Page transition, which is dual to the algebra localized on one exterior region of the bulk Schwarzschild-$AdS$ spacetime. When the CFT Hamiltonian is included, this boundary algebra transitions to type \Rmnum{2}${}_\infty$ \cite{Witten:2021unn,Chandrasekaran:2022eqq}. In the bulk, this corresponds to allowing fluctuations of the relative time-shift modes between the two asymptotic boundaries and subsequently \textit{gauging}\footnote{In this work, “gauging” refers to imposing the gravitational constraints associated with the symmetry under discussion.} the Killing boost symmetry—identified with the modular flow of the vacuum state \cite{Bisognano:1975ih,Sewell:1982zz,Sorce:2023gio}. The resulting boost-invariant subalgebra coincides with the modular crossed-product construction \cite{van1978continuous,takesaki2003theory}.  Being of type \Rmnum{2},  it admits a trace and therefore supports a well-defined notion of von Neumann entropy. For semiclassical states, the algebraic entropy matches the generalized entropy. In this picture, the time-shift modes—dual to the black hole energy—collectively form a clock that provides a reference frame for the boost symmetry, enabling the relational definition of boost-invariant operators.

Although highly instructive, the AdS/CFT discussion cannot be directly applied to de Sitter (dS) spacetime, where the situation is fundamentally different. In de Sitter spacetime, the absence of an asymptotic boundary  implies that there is no canonical choice of a static patch; instead, each timelike geodesic defines its own static patch and the corresponding cosmological horizon. This ambiguity directly  obstructs the precise definition of a local algebra, as there is no preferred region to which it can be associated. Furthermore, in the absence of a natural time-shift mode to serve as a clock, imposing boost invariance alone would lead to a trivial algebra.

Chandrasekaran, Longo, Penington and Witten (CLPW) \cite{Chandrasekaran:2022cip} made  a major  advance in defining local observable algebras in dS. The model they proposed  features an observer moving along a fixed geodesic and carrying a clock. This preferred geodesic selects a specific static patch, and it is natural to propose that the observer measures quantum fields along the geodesic. By the timelike tube theorem \cite{Witten:2023qsv,Borchers1961,Araki1963AGO,Strohmaier:2023hhy,Strohmaier:2023opz}, these measurements generate the full algebra of observables in the static patch.\footnote{The timelike tube theorem applies to the "timelike envelop"—the set of all events reachable by deforming the geodesic while keeping its endpoints fixed and maintaining its timelike character. For an observer on a geodesic, this region coincides with the associated static patch.} The clock  provides a reference frame for the boost-like symmetry that generates time translations within the static patch coordinates—identified as the modular flow of the Bunch-Davies vacuum. This yields a modular crossed product algebra, making the discussion of entropy tractable. Notably, by imposing a lower bound on the clock Hamiltonian's spectrum, the algebra of observables is  reduced  to type \Rmnum{2}${}_1$, in which the vacuum state attains the maximum entropy, consistent with the expectation that empty de Sitter space maximizes the generalized entropy \cite{Bousso:2000nf,Bousso:2000md}.

The success of obtaining a well-defined entropy in gravitational systems by imposing boost invariance has inspired a range of extensions to more general spacetimes \cite{Witten:2023qsv,Witten:2025xuc,Jensen:2023yxy,Kudler-Flam:2023qfl,Kudler-Flam:2024psh,AliAhmad:2023etg,Gomez:2023upk,Gomez:2023wrq,AliAhmad:2025ukh,Bahiru:2025ujp,Cao:2025els,Klinger:2025tvg,Chandrasekaran:2026pnc,Klinger:2026tws,Aguilar-Gutierrez:2023odp,Blommaert:2026lvp,Freidel:2026stu,Zhong:2026udo,Chandrasekaran:2026gvk}. These developments have led to significant conceptual advances, including an improved proof of the generalized second law (GSL) for semiclassical states \cite{Faulkner:2024gst}, as well as new insights into the Bekenstein bound and quantum null energy condition\cite{Kudler-Flam:2023hkl}. In parallel, the role of observers in gravitational systems has been explored in a variety of complementary frameworks. These include the constructions where the observer emerges intrinsically from quantum fields in slow-roll inflation \cite{Chen:2024rpx,Speranza:2025joj}, the models that treat the observer as a fully-quantized relativistic particle with a clock \cite{Kolchmeyer:2024fly}, and the descriptions in which the observer is represented by a Goldstone vector field \cite{Geng:2024dbl,Geng:2025bcb}. Related ideas have also been formalized using the framework of quantum reference frames (QRF) \cite{Fewster:2024pur,AliAhmad:2024vdw,AliAhmad:2024wja,DeVuyst:2024fxc}, which naturally encodes the insight that gravitational entropy is observer-dependent \cite{DeVuyst:2024fxc,DeVuyst:2024khu}.

More  recently, Kirklin \cite{Kirklin:2024gyl} proposed an interesting generalization beyond boost invariance. While a boost generates the additive group $\R$, Kirklin considered additional invariance under null translation in asymptotically AdS or flat black hole spacetimes. This leads to an enlarged two-dimensional group acting on the null coordinate $v$ along the horizon as
\begin{equation}
    G=\{v\to e^{2\pi t}v+s|t,s\in\R\}.
\end{equation}
To address the absence of a preferred notion of fixed exterior regions, Kirklin introduced the notion of \textit{dynamical cuts} of the horizon. These encode a relational definition of the exterior region, with the location of the cut promoted to a quantum degree of freedom transforming under null translations, furnishing a QRF for null translations. This construction allows a nontrivial gauge-invariant subalgebra and leads to a modified GSL valid beyond semiclassical states.

This progression from boosts to null translations raises a fundamental question: which second-order constraints should be imposed in the $G_N\to0$ linearized theory, and whether this procedure admits a natural endpoint, given the infinite-dimensional nature of the diffeomorphism group.  In spatially closed spacetimes with Killing horizons, such as the de Sitter space considered here, the restriction to the de Sitter Killing algebra is not an arbitrary truncation of the diffeomorphism group, but by linearization instabilities \cite{fischer1980structure,Arms:1982ea,Marsden_lectures,Arms:1986vk,Deser:1973zza,Moncrief1,Moncrief:1976un}. The ordinary first-order linearized diffeomorphism constraints are taken to be part of the construction of the perturbative QFT algebra $\A_{QFT}$, for instance through dressing to construct gauge-invariant operators \cite{Donnelly:2015hta,Donnelly:2016rvo,Giddings:2018umg,Giddings:2022hba,Giddings:2025xym,Giddings:2025bkp,Francois:2024rdm,Francois:2025shu,Francois:2025lqn}. What remains is a finite set of additional integrability conditions: for each background Killing field, the corresponding linearized constraint vanishes identically, while the first non-trivial generator appears at second order. Classically, a linearized solution is integrable to an exact solution only if all such second-order Killing generators vanish \cite{DeVuyst:2024grw}, which reflects the consistency of the perturbative expansion. Quantum mechanically, strong arguments suggest the necessity and sufficiency of imposing quantum Killing constraints to avoid inconsistent solutions \cite{Moncrief:1978te,Moncrief:1979bg,Higuchi:1991tk,Higuchi:1991tm,Losic:2006ht}, thus enforcing the corresponding symmetries as gauge symmetries.\footnote{For spacetimes with an asymptotic boundary or without Killing symmetries, no instabilities arise. Imposing the constraints depends on whether or not one is interested in a limited number of certain second-order observables, see the discussion in \cite{DeVuyst:2024grw}.} 

Consequently, in the de Sitter case, consistency of the perturbative expansion singles out the finite set of second-order constraints associated with the $SO(1,d)$ isometry group. In this sense, the CLPW framework implements only a subset of these constraints. The CLPW  construction relies on a \textit{fixed} geodesic, which does not transform covariantly under dS isometries. The isometry group is then required to preserve both the metric and this geodesic. Therefore, the presence of a single observer  reduces the symmetry to the subgroup
\begin{equation}
    SO(1,d)\to\R\times SO(d-1).
\end{equation}
The residual symmetry is then gauged by equipping the observer with a clock and an orthogonal frame, thereby constructing a nontrivial gauge-invariant algebra. This approach, however, is incomplete from the perspective of full gravitational constraints. Crucially, in a spacetime where all boost directions are physically equivalent, no single direction should be preferred \textit{a priori}. 

A limitation of the CLPW framework becomes apparent when considering multiple observers. The original construction already requires an antipodal observer to ensure a consistent representation of the gauge-invariant subalgebra on the physical Hilbert space.\footnote{Subsequent studies show that this representation is faithful (and hence a von Neumann algebra) if and only if there exists another observer on the opposite patch \cite{DeVuyst:2024fxc,DeVuyst:2024khu}. A physical explanation may involve the classical impossibility of deforming a single static patch without also deforming the opposite patch—assuming spherical symmetry and the weak energy condition \cite{folkestad2024subregion}—or the quantum mechanical fact that a single-particle state in dS fails to be annihilated by all dS Killing generators \cite{Kaplan:2024xyk}.}  Nevertheless, within the CLPW setup, maintaining boost invariance as required for a well-defined entropy forces any additional observers to be coincident or strictly antipodal. Although one may consider the observers on non-geodesic trajectories (e.g., general boost orbits in \cite{DeVuyst:2024fxc,DeVuyst:2024khu}), it nonetheless precludes two separate free-falling observers on generic, non-antipodal geodesics. Any framework intended to describe local physics in de Sitter space should accommodate observers on arbitrary geodesics—a feature that is not captured by the original CLPW construction.

In short, a satisfying scenario must simultaneously achieve two requirements: gauging the full $SO(1,d)$ symmetry while incorporating multiple observers along arbitrary geodesics. These two requirements appear incompatible within the CLPW framework. 

To resolve this tension, we introduce the notion of "covariant observer"\footnote{In standard terminology, a "dynamical reference frame" typically refers to a classical frame, distinct from quantum reference frame. Although the observer in our model might be called a "dynamical observer", the term has already been used by Kolchmeyer and Liu in \cite{Kolchmeyer:2024fly}. We therefore adopt the term "covariant observer" to emphasize the fact that the geodesic itself transforms covariantly.\label{notion}}. The central idea is to promote the observer’s geodesic from a fixed object to a dynamical one, whose defining data are treated as quantum variables. In particular, while the observer is still described by a timelike geodesic, the overall positions of this geodesic in spacetime—characterized by its conserved charges under the de Sitter isometry group—is no longer fixed, but is subject to quantum fluctuations.  This embodies two essential and novel principles:
\begin{enumerate}
    \item \textbf{Dynamical geodesic}: The observer's trajectory is not a preferred path singled out by hand, but a dynamical entity whose specification \textit{transforms covariantly} under the full de Sitter isometry group, thereby preserving the full spacetime symmetry. This ensures that no particular static patch is privileged, and is conceptually natural for a complete theory of quantum gravity, which should contain no external, non-dynamical elements. When constructed intrinsically, the observer must transform under isometry in the same manner as any other physical degree of freedom.
    \item \textbf{Quantum superposition of geodesics}: The classical ambiguity in selecting a preferred static patch in de Sitter space—arising from the absence of an asymptotic boundary—is resolved at the quantum level. Rather than fixing a single geodesic, we allow the observer to be described by a quantum superposition of classical geodesics related by isometries. Equivalently, the associated static patch becomes a fluctuating notion, reflecting the fact that all timelike geodesics in de Sitter space are physically equivalent and should be treated democratically in a quantum theory.
\end{enumerate}
Consequently, the covariant observer naturally provides a reference frame for the entire $SO(1,d)$ group. Upon quantization, this gives rise to a Hilbert space $L^2(SO(1,d))$, which supports a unitary representation of the isometry group and serves as a quantum reference frame.

Although our proposal of a covariant observer is formulated at the kinematical level, we introduce a simple dynamical model to clarify the physical origin of the fluctuating geodesic. Classically, a timelike geodesic in de Sitter space is completely characterized by its conserved charges associated with the isometry group. We therefore consider an action in which the observer carries a set of conserved charges, each one conjugate to a generator of de Sitter isometry group:
\begin{equation}
      S=\int d\tau (\dot p_A q^A-q^A\xi_A^\mu\dot x_\mu-m),
\end{equation}
so that the choice of geodesic is encoded in the integrals of motion rather than fixed background data. The fluctuations of the observer’s geodesic thus originate from the quantization of these charges, which is naturally understood as a quantum superposition of distinct classical geodesics labeled by their values, rather than fluctuations around a single worldline.

Inspired by the timelike tube theorem, we propose that the algebra accessible to an observer is generated by all degrees of freedom within its fluctuating static patch. The corresponding kinematical algebra should not be viewed as a fixed static-patch algebra $\A_{QFT}(0,0)$ tensored with an independently appended quantum reference frame $B(L^2(SO(1,d)))$.  Rather, the observer state selects the geodesic, and the QFT algebra associated with the corresponding static patch. Thus, the local algebra is defined relationally with respect to the observer's fluctuating geodesic, rather than with respect to a fixed background region. From the perspective of a given observer, the other observers can be viewed as part of the matter system, and the algebra therefore also includes the degrees of freedom of any other observers entering the patch. Since the geodesic fluctuates, the number of degrees of freedom within a given static patch also fluctuates in the presence of multiple observers, naturally capturing their entry and exit. Imposing the $SO(1,d)$ gauge constraints then yields a nontrivial $dS$-invariant algebra, which can be understood as an average over all static patches and geodesic configurations. In this way, our framework generalizes the notion of a local algebra for a fixed region to that of a \textit{fluctuating region}.\footnote{Following the terminological discussion in footnote \ref{notion}, one might call this a \textit{dynamical region}.} 

Remarkably, although this construction  involves degrees of freedom distributed over a full Cauchy surface, the resulting algebra is of type \Rmnum{2}, fundamentally due to its entanglement with the algebra of the fluctuating causal complement region, which is also of type \Rmnum{2}. This  behavior is in sharp contrast with the type \Rmnum{1} algebra found in a similar approach \cite{Kolchmeyer:2024fly}, and a detailed comparison is provided at the end of the paper. 

The fluctuating nature of the observer’s static patch has important implications in the first law for entropy. From the perspective of a given observer, the presence of an event horizon implies that an entropy should be associated with inaccessible degrees of freedom. However, when both the location of the static patch and the number of internal degrees of freedom fluctuate, standard formulations of the first law for a fixed subregion \cite{Gibbons:1977mu,Jensen:2023yxy} are not directly applicable. The law is nevertheless indispensable for identifying algebraic entropy with generalized entropy for semiclassical states.  This necessitates a quantum generalization of the first law for fluctuating regions with variable internal degrees of freedom. We find that such a generalization remains possible, relying on the fact that the expectation value of the area operator exhibits a linear behavior under superpositions of states, allowing a controlled extension of the first law. 

Notably, within the algebra associated with a single covariant observer, the observer’s own degrees of freedom and those of other observers—treated as part of the matter system—contribute in inequivalent ways. This asymmetry provides a concrete realization of subsystem relativity \cite{AliAhmad:2021adn,Castro-Ruiz:2021vnq,delaHamette:2021oex,Hoehn:2023ehz}, and supports the view that gravitational entropy is inherently observer-dependent \cite{DeVuyst:2024fxc,DeVuyst:2024khu}.

For simplicity,  we first develop our model in $dS_2$ space, postponing the generalization to higher dimensions to a later sketch.  One  technical advantage of $dS_2$ is that, in the embedding space $\R^{1,2}$, any timelike geodesic can be identified as the intersection of the $dS_2$ with a plane through the origin orthogonal to a spacelike unit vector. This characterization greatly facilitates the description of how geodesics transform under the isometry group and the analysis of their mutual causal relationships—both essential in constructing the observer's algebra.

The paper is organized as follows. In section \ref{sec:classical}, we perform a classical analysis of timelike geodesics in $dS_2$ and their transformations under $SO(1,2)$, whose group parameter space is $(\phi,s,t)$. We demonstrate that timelike geodesics can be parameterized by $(\phi,s)$, and points on the geodesics by $(\phi,s,t)$ via their transformation laws, thus establishing  a classical reference frame. Furthermore, as a prerequisite for the subsequent algebraic construction, we derive the necessary and sufficient condition for one geodesic to be causally connected to a segment of another, and explicitly identify that segment. 

In section \ref{sec:algbera}, we construct the algebra for a covariant observer $\O$. We begin by quantizing  the classical parameter space $(\phi,s,t)$ to form the Hilbert space $\HH_\O=L^2({SO(1,2)})$ that serves as a QRF, with the state $\ket{\phi,s,t}_\O$ describing an observer localized at the point specified by $(\phi,s,t)$. Inspired by the timelike tube theorem, we propose that an observer can access QFT modes within the associated static patch, and is equipped with a clock and a boost Hamiltonian to measure time and evolve along its geodesic. We then introduce an action that model a covariant observer,  in which the observer’s geodesic is treated as a quantum degree of freedom through the quantization of the conserved charges associated with de Sitter isometries. When multiple covariant observers $\O_a,\O_b,\cdots$ are present, we propose that $\O_a$ can access the algebraic degrees of freedom of $\O_b$ associated with the portion of the latter's trajectory that lies within $\O_a$'s static patch. The resulting $dS$-invariant subalgebra is an averaged version of the modular crossed product algebra, shown to be of type \Rmnum{2} by constructing a faithful normal trace. We also construct the physical Hilbert space via group averaging and identify the algebra’s representation, finding it faithful if and only if at least one other covariant observer is present. 

Section \ref{sec:entropy} is devoted to the study of entropy of semiclassical states.  By imposing a UV cutoff in QFT, and proposing a sensible quantum generalization of the first law for fluctuating region along with a fluctuating number of internal degrees of freedom, we demonstrate that the algebraic entropy coincides with the generalized entropy. Our methodology in this section systematically follows the procedure well-established in \cite{DeVuyst:2024fxc,DeVuyst:2024khu}, closely mirroring the steps detailed in \cite{Kirklin:2024gyl}. 

In section \ref{sec:higher dimension} we outline the generalization to higher-dimensional $dS_d$ spacetime. In this case,  the symmetry group requires  that an observer carries not only a clock but also an orthogonal frame to constitute a complete QRF. We extend our previous measurement proposals accordingly, to that an observer can now also access the orientation of other observers' orthogonal frames within its own static patch.

We leave section \ref{conclusion} for an outline of our main results, several promising directions, and  comparison with other existing works relevant to our study.

\section{Geodesics  in $dS_2$ \label{sec:classical}}
This section establishes the classical groundwork for the covariant observer by analyzing timelike geodesics in two-dimensional de Sitter spacetime $dS_2$, their transformations under the $SO(1,2)$ isometry group and their causal relationships—all essential for the subsequent algebraic construction.

The $dS_2$ spacetime is defined as the hypersurface
\begin{equation}
    -X_0^2+X_1^2+X_2^2=1
\end{equation}
embedded in $\R^{1,2}$. The isometry group $SO(1,2)$, induced from the Lorentz group of $\R^{1,2}$, has generators:  $B_1$ and $B_2$ for the boosts in the $X^0-X^1$ and $X^0-X^2$ plane, respectively, and $R$ for the rotation in the $X^1-X^2$ plane, with the commutation relations
\begin{equation}\label{eq:commuation-killing}
    [R,B_1]=-B_2,\quad[R,B_2]=B_1,\quad[B_1,B_2]=R.
\end{equation}
We adopt the following parametrization of an element in $SO(1,2)$:
\begin{equation}\label{eq:isometry element}
    \begin{aligned}
        g(\phi,s,t)&=e^{-\phi R}e^{-s B_2}e^{-t B_1}
    \end{aligned}
\end{equation}
with $t,s\in\R,$ and $ \phi\in[-\pi,\pi]$, which corresponds to performing a boost in the $X^0-X^1$ plane, followed by a boost in the $X^0-X^2$ plane, and then a rotation in the $X^1-X^2$ plane. This choice facilitates the description of transformations between geodesics. The left/right-invariant Haar measure in these coordinates are
\begin{equation}\label{eq:group measure}
    d\mu_L=d\mu_R=\cosh s\,d\phi dsdt.
\end{equation}
Although the explicit form of the group multiplication is complicated, we denote it symbolically as
\begin{equation}
    g(\phi_1,s_1,t_1)g(\phi_2,s_2,t_2)=g(\phi_{1+2},s_{1+2},t_{1+2}),\quad g(\phi,s,t)^{-1}=g(\phi_{-},s_{-},t_{-}).
\end{equation}
It should be emphasized that for general group elements
\begin{equation}
    (\phi_{1+2},s_{1+2},t_{1+2})\neq(\phi_1+\phi_2,s_1+s_2,t_1+t_2)
\end{equation}
and similarly for the inverse element.

\subsection{Isometry action and reference frames }
We begin with a timelike geodesic $L_{0}=(\sinh\tau,\cosh\tau,0)$.   Under the action of an isometry $\phi_g$ associated with a given $g\in SO(1,2)$, it transforms covariantly as:
\begin{equation}
  \phi_g:  L_{0}\to L_{0}'=\phi_g(L_{0}).
\end{equation}
Notably, a boost in the $X^0-X^1$ plane preserves $L_0$ with a translation along it: 
\begin{equation}
    \tau\to\tau+t.
\end{equation}
In general, the group action mixes transformations between geodesics and translations along them.
We now show that the points on timelike geodesics form a complete reference frame for $SO(1,2)$, as the group action on them uniquely specifies the isometry. 

\begin{figure}[htbp]
\centering
\includegraphics[width=.4\textwidth]{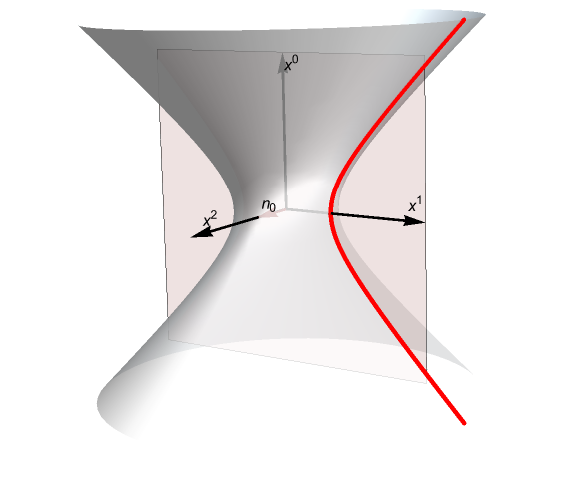}
\caption{The embedding of $dS_2$ in $\R^{1,2}$ is depicted, with the timelike geodesic $L_{0}=(\sinh\tau,\cosh\tau,0)$ shown in red. This geodesic is the intersection of $dS_2$ with the $X^0-X^1$ plane, which is orthogonal to the spacelike unit vector $\hat{n}_0=(0,0,1)$.  \label{fig:1}}
\end{figure}

A useful geometric description arises from observing that $L_{0}=(\sinh\tau,\cosh\tau,0)$ is the intersection of the $dS_2$ with the plane through the origin and orthogonal to the spacelike unit vector  $\hat{n}_0=(0,0,1)$. In general, such a plane intersects $dS_2$ in two disconnected timelike geodesics. To uniquely select a branch,  we demand that the vector $\epsilon^{\mu\nu\rho}v_{\nu}(n_0)_{\rho}$ is future-directed, where $\hat{v}=(0,1,0)$ fix the point on $L_0$ at $\tau=0$ and $\epsilon^{\mu\nu\rho}$ denotes the Levi-Civita symbol. See figure.~\ref{fig:1}. We therefore say that $L_0$ is uniquely determined by $\hat{n}_0$. 
This construction can be generalized as follows: given any spacelike unit vector $\hat{n}$, there exists a timelike unit vector $\hat{u}$ and a spacelike unit vector $\hat{v}$ such that
\begin{equation}
    \hat{u}\cdot \hat{v}=\hat{u}\cdot \hat{n}=\hat{v}\cdot \hat{n}=0
\end{equation}
with  $\epsilon^{\mu\nu\rho}v_{\nu}n_{\rho}$ being future-directed to select a branch. This defines a timelike geodesic:
\begin{equation}\label{eq:geodesic}
    L_{\hat{n}}=\hat{u}\sinh\tau+\hat{v}\cosh \tau
\end{equation}
which represents the intersection of $dS_2$ with the plane through the origin and orthogonal to $\hat{n}$.
Thus, any timelike geodesic is parametrized by a spacelike unit vector $\hat{n}$,  and points on it are labeled by $(\hat{n},\tau)$, fixing $\tau=0$ on the $X^1-X^2$ plane. Therefore, the question of how points on timelike geodesics transform under $SO(1,2)$ is translated into how $(\hat{n},\tau)$ changes.

Recall that our parametrization of $SO(1,2)$ separates the transformations that preserve $L_0$ (translations along it) from those that change $L_0$ itself. We find that under $g(\phi,s,t)$, induced from the Lorentz transformation in $\R^{1,2}$, the reference frame transforms as:
\begin{equation}\label{eq:geodesic transformed}
   \phi_g: \hn_0\to \hn(\phi,s)=(\sinh s,-\cosh s\sin\phi,\cosh s\cos\phi),\quad \tau=0\to\tau=t
\end{equation}
This establishes a bijection between the group parameters $(\phi,s,t)$ and the transformed reference frame $(\hn,\tau)$. We therefore introduce the notation $L_{(\phi,s)}$ for the geodesic determined by $\hn(\phi,s)$ and $P_{(\phi,s,t)}$ for the point at proper time $\tau=t$ on it. We also let $\M_{(\phi,s)}$ denote the causal region, or equivalently the static patch, associated with the geodesic $L_{(\phi,s)}$.  The group acts naturally on the geodesic and corresponding causal region by definition:
\begin{equation}
    \phi_{g(\phi_1,s_1,t_1)}L_{(\phi_2,s_2)}=L_{(\phi_{1+2},s_{1+2})},\quad \phi_{g(\phi_1,s_1,t_1)}P_{(\phi_2,s_2,t_2)}=P_{(\phi_{1+2},s_{1+2},t_{1+2})}
\end{equation}
This demonstrates that the space of points $P_{(\phi,s,t)}$ on timelike geodesics provides a complete reference frame for the isometry group $SO(1,2)$.

Finally, the generator of translation along  $L_{(\phi,s)}$ is given by
\begin{equation}
  B_{(\phi,s)}= B_1\cos\phi\cosh s+B_2\sin\phi\cosh s+R\sinh s
\end{equation}
which by construction satisfies 
\begin{equation}
    \phi_{e^{-t'B_{(\phi,s)}}}P_{(\phi,s,t)}=P_{(\phi,s,t+t')}
\end{equation}

\subsection{Causal contact between geodesics}
To construct the algebra for multiple observers, we must first determine the conditions under which one observer can causally access another. This subsection derives the necessary and sufficient condition for a segment of a geodesic $L_{(\phi_2,s_2)}$ to lie within the static patch of another geodesic $L_{(\phi_1,s_1)}$.
\begin{figure}[htbp]
\centering
\includegraphics[width=.4\textwidth]{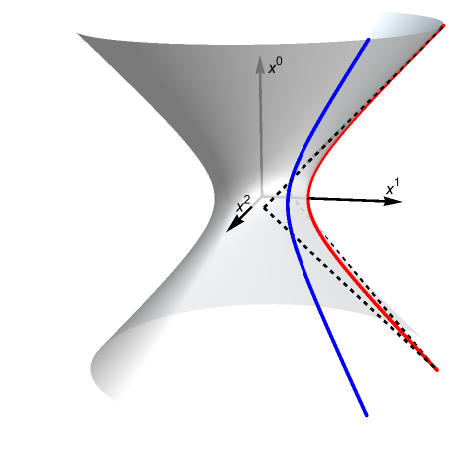}
\caption{Causal contact between geodesics in $dS_2$. The timelike geodesic $L_0$ (red) and the boundary of its static patch $\M_0$ (black dashed)
 are shown, together with another geodesic $L_1=(\sinh \tau,\cos\frac{\pi}{4}\cosh\tau,\cos\frac{\pi}{4}\cosh\tau)$ (blue) that partially lies within $\M_0$.  \label{fig:2}}
\end{figure}

We begin with a criterion for a point to be inside a static patch. On $dS_2$, the causal nature of the geodesic connecting two points $X$ and $X'$ is determined by their inner product $P(X,X')=X\cdot X'$ in the embedding space: the connecting geodesic is timelike if $P(X,X')>1$, null if $P(X,X')=1$ and spacelike if $0<P(X,X')<1$.
Similar results hold for $X$ and $-X'$ if $P(X,X')<0$. A point $X=(a,b,c)$ lies within the static patch $\M_0$ of the reference geodesic $L_0$ if and only if it is causally connected to both future and past infinity of $L_0$. This translates to the condition:
\begin{equation}
    \left\{
    \begin{aligned}
        &\lim_{\tau\to\infty }-a\sinh\tau+b\cosh\tau\geq1  \\
        &\lim_{\tau\to-\infty }-a\sinh\tau+b\cosh\tau\geq1
    \end{aligned}
    \right.
\end{equation}
which simplifies to the equivalent condition:
\begin{equation}\label{eq:point locate}
    0\leq|a|\leq b.
\end{equation}

Now consider a general timelike geodesic $L_{(\phi,s)}$. The associated orthonormal frame vectors are given by:
\begin{equation}
    u^\mu=(\cosh s,-\sinh s\sin\phi,\sinh s\cos\phi),\quad v^\nu=(0,\cos\phi,\sin\phi).
\end{equation}
Substituting the parametrization \eqref{eq:geodesic} for points on $L_{\phi,s}$ into the condition \eqref{eq:point locate}, we find that the portion of the geodesic inside $\M_0$ must satisfy:
\begin{equation}
    |\sinh\tau\cosh s|\leq\cosh\tau\cos\phi-\sinh\tau\sinh s\sin\phi.
\end{equation}
A solution for $\tau$ exists if and only if 
\begin{equation}
    \cos\phi\geq0\quad\mathrm{i.e.}\quad\phi\in[-\frac{\pi}{2},\frac{\pi}{2}].
\end{equation}
When this condition is met, the accessible segment in terms of proper time is: 
\begin{equation}
    \tanh\tau\in\left[-\frac{\cos\phi}{\cosh s-\sinh s\sin\phi},\frac{\cos\phi}{\cosh s+\sinh s\sin\phi}\right].
\end{equation}

Since the isometries preserve causal structure, we can generalize this result to arbitrary pairs of geodesics by applying an appropriate isometry. Define relative group parameters $(\phi_{-1+2},s_{-1+2},t_{-1+2})$ through group composition as $g^{-1}(\phi_1,s_1,t_1)g(\phi_2,s_2,t_2)=g(\phi_{-1+2},s_{-1+2},t_{-1+2})$, then a segment of the geodesic $L_{(\phi_2,s_2)}$  is visible to $L_{(\phi_1,s_1)}$ if and only if $\phi_{-1+2}$ satisfies
\begin{equation}\label{eq:geodesic-seen}
    \phi_{-1+2}\in[-\pi/2, \pi/2].
\end{equation}
As the proper time parameter along $L_{(\phi,s)}$ is identified with the group parameter $t$, the accessible segment is given by:
\begin{equation}\label{eq:geodesic-seen-part}
  \tanh t \in\left[-\frac{\cos\phi_{-1+2}}{\cosh s_{-1+2}-\sinh s_{-1+2}\sin\phi_{-1+2}},\frac{\cos\phi_{-1+2}}{\cosh s_{-1+2}+\sinh s_{-1+2}\sin\phi_{-1+2}}\right].
\end{equation}
For brevity in subsequent sections, we denote this accessible proper time interval as
\begin{equation}
    t\in [F_{-1+2},G_{-1+2}].
\end{equation}
where $F_{-1+2}$ and $G_{-1+2}$ are functions of the relative group parameters $\phi_{-1+2}$ and $s_{-1+2}$, as defined in \eqref{eq:geodesic-seen-part}. 

\section{Algebra for covariant observers}\label{sec:algbera}

We now impose the second-order gravitational constraints associated with the full de Sitter isometry group $SO(1,2)$ and implement the notion of a covariant observer introduced in the Introduction. The purpose of this section is to provide an explicit algebraic realization of a single covariant observer and to construct the corresponding gauge-invariant observable algebra.

Classically, the observer is characterized by a timelike geodesic transforming covariantly under $SO(1,2)$. Upon quantization, its degrees of freedom furnish a quantum reference frame for the isometry group. A simple dynamical model realizing the covariant observer, in which the geodesic naturally exhibits quantum fluctuations, will be introduced later in this section. At a given quantum state of the observer, we postulate that the accessible algebra is generated by all quantum field degrees of freedom contained within the associated static patch, as motivated by the timelike tube theorem. Consistent with the principle that other observers are treated as part of the matter system, the algebra also includes the degrees of freedom of any other observers entering the patch, according to the classical causal criteria derived in \ref{sec:classical}. Imposing the gravitational constraints then yields a gauge-invariant subalgebra, which takes the form of an "averaged" modular crossed product. This algebra admits a faithful normal trace and is therefore type \Rmnum{2}. This section follows a method similar to ref.~\cite{Kirklin:2024gyl}.

Before constructing the algebra, let us recall the perturbative status of the gravitational constraints relevant for the present work. In the $G_N\to 0$ theory around a fixed de Sitter background, the ordinary linearized diffeomorphism constraints are first-order constraints acting on the graviton perturbation. We assume that these constraints, together with possible matter gauge constraints, have already been implemented in defining the perturbative QFT algebra $\A_{\rm QFT}$, which is built from observables invariant under linearized diffeomorphisms, as can be achieved by dressing methods \cite{Donnelly:2015hta,Donnelly:2016rvo,Giddings:2018umg,Giddings:2022hba,Giddings:2025xym,Giddings:2025bkp,Francois:2024rdm,Francois:2025shu,Francois:2025lqn}.  

Background Killing fields have a different status. Since they leave the background invariant, the bulk contribution of the corresponding linearized constraint vanishes; equivalently, the integrated constraint reduces to a boundary term, which is absent in spatially closed de Sitter space. The first non-trivial generator associated with such a symmetry therefore appears at second order. The linearization stability conditions then requires the finite set of charges, one for each independent de Sitter Killing field, to vanish. It is these second-order Killing constraints, rather than arbitrary second-order diffeomorphism constraints, that will be imposed below.  For a detailed discussion, see Ref. \cite{DeVuyst:2024grw}.

With this understood, we take QFT algebra $\A_{QFT}(\phi,s)$, associated with the static patch $\M(\phi,s)$, to be the dressed linearized QFT algebra acting on the Hilbert space $\HH_{QFT}$. The de Sitter isometry group is unitarily represented on $\HH_{QFT}$. We denote the corresponding quantum Killing generators by $H_1,H_2,J$, with corresponding second-order constraints imposed on the combined QFT/observer system. They satisfy the commutation relations inherited from the Killing algebra \eqref{eq:commuation-killing}:
\begin{equation}
    [J,H_1]=iH_2,\quad[J,H_2]=-iH_1,\quad[H_1,H_2]=-iJ.
\end{equation}
A general group element is parameterized according to the classical parametrization \eqref{eq:isometry element} as
\begin{equation}\label{eq:parametrise-QFT}
    U_{QFT}(\phi,s,t)=e^{-i\phi J}e^{-is H_2}e^{-itH_1}
\end{equation}

\subsection{Algebra for a single covariant observer}
The classical parameters $(\phi,s,t)$, which label points on timelike geodesics, constitute a natural reference frame for $SO(1,2)$. We quantize this space by taking the Hilbert space to be $\HH_{\O}=L^2(SO(1,2))$ with orthonormal basis states $\ket{\phi,s,t}$ satisfying
\begin{equation}\label{eq:ortho-complete}
    \langle\phi_1,s_1,t_1|\phi_2,s_2,t_2\rangle_\O=\delta(\phi_{-1+2})\delta(s_{-1+2})\delta(t_{-1+2}),\quad\mathbf{1}=\int d\phi ds dt\ket{\phi,s,t}_\O\bra{\phi,s,t}_\O .
\end{equation}
We denote the generators  of $SO(1,2)$ on $\HH_{\O}$ as $\hat{l},\hat{d}_1,\hat{d}_2$ accordingly, which satisfy the commutation relations:
\begin{equation}
    [\hat{l},\hat{d}_1]=i\hat{d}_2,\quad[\hat{l},\hat{d}_2]=-i\hat{d}_1,\quad[\hat{d}_1,\hat{d}_2]=-i\hat{l}.
\end{equation}
Then a group element acts unitarily on $\HH_{\O}$ via
\begin{equation}\label{eq:action on QRF}
    U_{\O}(\phi_1,s_1,t_1)\ket{\phi_2,s_2,t_2}_{\O}=\sqrt{\frac{\cosh s_2}{\cosh s_{1+2}}}\ket{\phi_{1+2},s_{1+2},t_{1+2}}_\O
\end{equation}
where $U_\O(\phi,s,t)=e^{-i\phi\hat{l}}e^{-is\hat{d}_2}e^{-it\hat{d}_1}\in SO(1,2)$. The prefactor  $\sqrt{\frac{\cosh s_2}{\cosh s_{1+2}}}$ ensures the unitarity of  $U_\O(\phi,s,t)$, as can be verified using the left-invariance of the Haar measure:
\begin{equation}
    \cosh s_2 d\phi_2ds_2dt_2=\cosh s_{1+2}d\phi_{1+2}ds_{1+2}dt_{1+2}
\end{equation}

A \textit{covariant observer} $\O$ is defined abstractly as a physical system with Hilbert space $\HH_\O=L^2(SO(1,2))$, along with the algebra of observables constructed below. The state $\ket{\phi,s,t}_\O$ describes an observer whose trajectory is the geodesic $L_{(\phi,s)}$, located at the specific point $P_{(\phi,s,t)}$ at which the QFT measurements are performed. Crucially, the observer's Hilbert space encompasses a superposition of all possible geodesics, reflecting the quantum fluctuation of the static patch itself. This framework provides an ideal quantum reference frame for $SO(1,2)$: states with different $(\phi,s,t)$ are orthogonal. 

Including the degrees of freedom of the observer $\O$, the kinematical Hilbert space is then a tensor product
\begin{equation}
    \HH_{kin}=\HH_{QFT}\otimes\HH_{\O}.
\end{equation}
The gauge generators of $SO(1,2)$ on $\HH_{kin}$ now include the contribution of QFT and the observer:
\begin{equation}\label{eq: constraint single}
    \c_1=H_1+\hat{d}_1,\quad \c_2=H_2+\hat{d}_2,\quad\d=J+\hat{l}
\end{equation}
We then parameterize the corresponding gauge group as:
\begin{equation}
      U(\phi,s,t)=e^{-i\phi \d}e^{-is \c_2}e^{-it\c_1}.
\end{equation}

The algebra of observables for $\O$ is constructed as follows. For an observer in state $\ket{\phi,s,t}_\O$, the accessible operators include:
\begin{enumerate}
    \item Its proper time $\hat{t}$.
    \item The generator of translations along its own geodesic, enabling motion along the trajectory,
\begin{equation}
    \hat{d}_{(\phi,s)}:=U_\O(\phi,s,0)\hat{d}_1 U^\dagger_\O(\phi_-,s_-,0)=\hat{d}_1\cos\phi\cosh s+\hat{d}_2\sin\phi\cosh s+\hat{l}\sinh s
\end{equation}
\item The QFT algebra $\A_{QFT}(\phi,s)$ within its static patch $\M_{(\phi,s)}$, motivated by the timelike tube theorem.
\end{enumerate}
The full kinematical algebra, denoted by $\A$, is therefore generated by
\begin{equation}
 \A_{QFT}(\phi,s)\otimes\{\hat{d}_{(\phi,s)}, \hat{t}\}''\ket{\phi,s,t}_\O\bra{\phi,s,t}_\O
\end{equation}
with $\forall s,t\in(-\infty,\infty)$ and $\forall\phi\in[-\pi,\pi]$. Here, we denote $\{\hat{d}_{(\phi,s)}, \hat{t}\}''$  as the von Neumann algebra generated by $\hat{d}_{(\phi,s)}$ and $\hat{t}$, being a type \Rmnum{1}${}_\infty$ factor. 
This expression should not be read as the tensor product of a fixed QFT algebra $\A_{QFT}(0,0)$ with an independently added QRF. Rather, the QFT algebra and the observer state are correlated: the component $\ket{\phi,s,t}_\O\bra{\phi,s,t}_\O$ is accompanied by the static-patch algebra $ \A_{QFT}(\phi,s)$. In this sense the observer does not merely provide external clock or frame variables for a pre-existing local algebra. The local algebra itself follows the quantum state of the observer's geodesic. The role of $\A_{QFT}(0,0)$ is therefore only to provide a fiducial representative from which the other static-patch algebras are obtained by de Sitter isometries.
We also note the center of the $\A$ is
\begin{equation}
    \Z(\A)=\{\hat{\phi},\hat{s}\}''.
\end{equation}

We now impose invariance under the gauge generators \eqref{eq: constraint single}. The operator of interest is of the form
\begin{equation}\label{eq:operator form}
   \begin{aligned}
        A=&\int d\phi_{1} ds_{1}dt_{1} a(\phi_{1},s_{1} ,t_{1})\otimes  e^{-if(\phi_{1},s_{1} ,t_{1})\hat{d}_{(\phi_{1},s_{1} )}}g(\hat{t})\ket{\phi_{1},s_{1} ,t_{1}}_\O\bra{\phi_{1},s_{1} ,t_{1}}_\O\\
        =&\int d\phi_{1} ds_{1}dt_{1} a(\phi_{1},s_{1} ,t_{1})\otimes g(t_1)\ket{\phi_{1},s_{1} ,t_{1}+f(\phi_{1},s_{1} ,t_{1})}_\O\bra{\phi_{1},s_{1} ,t_{1}}_\O
   \end{aligned}
\end{equation}
with $a(\phi,s,t)\in\A_{QFT}(\phi,s)$, and $f:SO(1,2)\to \R, g:\R\to\R$ being arbitrary smooth functions. Such operators are dense in the kinematical algebra $\A$. Using \eqref{eq:action on QRF} and the left-invariance of the Haar measure, one verifies that
\begin{equation}
    \begin{aligned}
        &U(\phi_{2},s_{2} ,t_{2})AU^\dagger(\phi_{2},s_{2} ,t_{2})\\
        =&\int d\phi_{1} ds_{1}dt_{1} U_{QFT}(\phi_{2},s_{2} ,t_{2})a(\phi_{-2+1},s_{-2+1} ,t_{-2+1})U_{QFT}^\dagger(\phi_{2},s_{2} ,t_{2})\\
        &\otimes g(t_{-2+1})\ket{\phi_{1},s_{1} ,t_{1}+f(\phi_{-2+1},s_{-2+1} ,t_{-2+1})}_\O\bra{\phi_{1},s_{1} ,t_{1}}_\O ~.
    \end{aligned}
\end{equation}
Then $A$ is invariant under the action of $SO(1,2)$ if and only if $f,g$ are constant functions, and
\begin{equation}
    U_{QFT}(\phi_{2},s_{2} ,t_{2})a(\phi_{-2+1},s_{-2+1} ,t_{-2+1})U_{QFT}^\dagger(\phi_{2},s_{2} ,t_{2})=a(\phi_{1},s_{1} ,t_{1}).
\end{equation}
Substituting this into \eqref{eq:operator form}, we obtain the following
\begin{equation}
    A=\int  d\phi dsdtU(\phi,s,t) \left(a \otimes  e^{-if\hat{d}_{1}}\right)U^\dagger(\phi,s,t)\ket{\phi,s,t}_\O\bra{\phi,s ,t}_\O
\end{equation}
where we denote $a:=a(0,0,0)$ and the constant $f:=f(0,0,0)$.

Let us introduce the dressing map $D$, which represents an average over the $SO(1,2)$ group, 
\begin{equation}
    D(a)=\int  d\phi dsdtU(\phi,s,t) aU^\dagger(\phi,s,t)\ket{\phi,s,t}_\O\bra{\phi,s ,t}_\O
\end{equation}
for $a\in \A_{QFT}(0,0)\otimes\{\hat{d}_1\}''$, with the properties that 
\begin{equation}
   D(ab)=D(a)D(b),~~~ D(a^\dagger)=D(a)^\dagger,~~~ D(D(a))=D(a). 
\end{equation} 
Under the map, $D(a)$ is $dS$-invariant, namely $D$ projects kinematical operators onto $dS$-invariant operators. Notice that $D(\hat{t})=0$ by construction. This leads to a full $SO(1,2)$-invariant subalgebra:\footnote{Our algebra differs in form from CLPW's, where the invariant algebra is $\{e^{ipH}ae^{-ipH},q|a\in\A\}''$, due to a different canonical convention:  our $\hat{t}$ is analogous to their $-p$ as $[q,p]=i$. The physical content is equivalent. Besides, let $\A\vee\B$ denote the von Neumann algebras generated by $\A$ and $\B$, one may want to schematically write $\A=\A_{QFT}(\hat{\phi},\hat{s})\vee\hat{d}_{(\hat{\phi},\hat{s})}\vee \hat{t}$ and $D(a)=U(\hat{\phi},\hat{s},\hat{t})aU^\dagger(\hat{\phi},\hat{s},\hat{t})$. Such formulas, however, do not apply since $\hat{d}_1$ fails to commute with $\hat{\phi}$, $\hat{s}$, $\hat{t}$.}
\begin{equation}\label{eq:algebra one inv}
    \A^G=D(\A_{QFT}(0,0)\otimes\{\hat{d}_1\}'')
\end{equation}
with trivial center. Then $D$ is an isomorphism from $\A_{QFT}(0,0)\otimes\{\hat{d}_1\}''$ to $\A^G$. If we insert $\delta(\phi)\delta(s)$ into the integrand of $D$, $D(\A_{QFT}(0,0)\otimes\{\hat{d}_1\}'')$ gives rise to the standard modular crossed product $e^{-i\hat{t}H_1}\A_{QFT}(0,0)e^{-i\hat{t}H_1}\rtimes \hat{d}_1$, as the QFT vacuum has modular operator $H_1$ \cite{Bisognano:1975ih,Sewell:1982zz,Witten:2018zxz}. Therefore, $\A^G$ can be interpreted as an averaged  modular crossed product of $\A_{QFT}(0,0)$ over the parameters $(\phi,s)$, corresponding to an average over all choices of reference timelike geodesics, or equivalently, static patches related by de Sitter isometries. 
Follows from the covariance of the family $\A_{QFT}(s,t)$ and the invariance of the Haar measure, this algebra is equivalent to
\begin{equation}
    \A^G=D(\A_{QFT}(\phi_0,s_0)\otimes\{\hat{d}_{(\phi_0,s_0)}\}'')
\end{equation}
for any $(\phi_0,s_0)$. Thus $\A_{QFT}(0,0)$ is only a coordinate choice in the description of the invariant algebra, without privileged role.

To verify the type \Rmnum{2}${}_\infty$ nature of $\A^G$, we now construct a specific trace functional. Recall that a key feature of the usual modular crossed product is that it converts a KMS state into a tracial state. Inspired by this, as we have the QFT vacuum $\Psi_{QFT}(\cdot)=\ev{\Omega}{\cdot}$ satisfies the KMS condition with respect to the modular operator $H_1$:
\begin{equation}\label{eq:KMS condition}
    \begin{aligned}
        \Psi_{QFT}(e^{iH_1\tau}Be^{-iH_1\tau}A)
        =\Psi_{QFT}(Ae^{iH_1(\tau+i)}Be^{-iH_1(\tau+i)}),
    \end{aligned}
\end{equation}
a trace functional $Tr:\A_a^G\to\C$ can then be defined as\footnote{A technical subtlety arises from the divergent inner product  $\langle0,0,t_a|0,0,t_a'\rangle_{\O_a}$, which renders the naive trace divergent. In practice, we use a renormalized version, effectively dividing by $\langle0,0|0,0\rangle_{\O_a}$. This just rescales the trace and makes no physical difference. One rigorous definition of trace can be got by defining the map $\gamma(a)\otimes\ket{0,0,t}_\O=a\ket{0,0,t}_\O$, similar to the treatment in ref.~\cite{Kirklin:2024gyl}. \label{divegence}}
\begin{equation}
    \begin{aligned}
        Tr(A)=&\Psi_{QFT}(\bra{0,0,0}_{\O}D(e^{-\frac{\hat{d}_{1}}{2}})AD(e^{-\frac{\hat{d}_{1}}{2}})\ket{0,0,0}_{\O})\\
        =&\Psi_{QFT}(\bra{0,0,0}_{\O}e^{-\frac{\hat{d}_{1}}{2}}Ae^{-\frac{\hat{d}_{1}}{2}}\ket{0,0,0}_{\O})
    \end{aligned}
\end{equation}
with $A\in\A^G$, and the second equality used the property $D(a)\ket{0,0,0}_\O=a\ket{0,0,0}_\O$ for $a\in \A_{QFT}(0,0)\otimes\{\hat{d}_1\}''$. This is a normal weight by construction. One may confirm that for a general elements in $\A^G$,
\begin{equation}\label{eq:operator inv}
    A=D(\int dt a(t)e^{-it\hat{d}_1})
\end{equation}
with $a\in\A_{QFT}(0,0)$, satisfying
\begin{equation}
    Tr(D(e^{\frac{\hat{d}_{1}}{2}})A^\dagger AD(e^{\frac{\hat{d}_{1}}{2}}))=\Psi_{QFT}(\int_{-\infty}^\infty a^\dagger(t)a(t))
\end{equation}
which shows $Tr$ is faithful. The cyclic property of $Tr$ is verified as follows. For any $A,B\in\A^G$, using their gauge invariance, one derives
\begin{equation}\label{eq:inv trans}
    \begin{aligned}
        \bra{\phi,s,t}_{\O} Ae^{-\frac{\hat{d}_{1}}{2}}\ket{0,0,0}_{\O}
        =U_{QFT}(\phi,s,t)e^{-\frac{H_1}{2}}\bra{0,0,0}_{\O} e^{-\frac{\hat{d}_{1}}{2}}A\ket{\phi_-,s_-,t_-}_{\O}e^{\frac{H_1}{2}}U^\dagger_{QFT}(\phi,s,t)
    \end{aligned}
\end{equation}
and then
\begin{equation}
    \begin{aligned}
& Tr(AB)\\
 =&\int_{-\infty}^\infty d\phi dsdt \Psi_{QFT}(\bra{0,0,0}_{\O}e^{-\frac{\hat{d}_{1}}{2}}A\ket{\phi,s,t}_{\O}\bra{\phi,s,t}_{\O}Be^{-\frac{\hat{d}_{1}}{2}}\ket{0,0,0}_{\O})\\
 =&\int_{\infty}^\infty  d\phi dsdt\Psi_{QFT}\big(\bra{\phi_-,s_-,t_-}_{\O}Ae^{-\frac{\hat{d}_{1}}{2}}\ket{0,0,0}_{\O}e^{-H_1}\bra{0,0,0}_{\O}e^{-\frac{\hat{d}_{1}}{2}}b\ket{\phi_-,s_-,t_-}_{\O}e^{H_1}\big)\\
 =&\int_{\infty}^\infty  d\phi dsdt\Psi_{QFT}(\bra{0,0,0}_{\O}e^{-\frac{\hat{d}_{1}}{2}}B\ket{\phi_-,s_-,t_-}_{\O}\bra{\phi_-,s_-,t_-}_{\O}Ae^{-\frac{\hat{d}_{1}}{2}}\ket{0,0,0}_{\O})\\
 =&Tr(BA)
    \end{aligned}
\end{equation}
where in the second equality we have used \eqref{eq:inv trans} and the fact that $\Psi_{QFT}$ is invariant under $U_{QFT}$, the third equality follows from the KMS condition \eqref{eq:KMS condition} with $\tau=0$. In summary, we obtain a faithful normal trace, with $Tr(1)=\infty$ due to the divergence in $\bra{0,0,0}_{\O}e^{-\hat{d}_1}\ket{0,0,0}_{\O}$ (we have ignored the divergence from $\langle0,0|0,0\rangle_{\O_a}$ as discussed in Footnote.~\ref{divegence}), which confirms that $\A^G$ is a type \Rmnum{2}${}_\infty$ factor.

Physically, the dressing $D$ represents an average over all static patches due to the quantum fluctuation of geodesic. Since each constituent algebra is type \Rmnum{2}, the final averaged algebra remains that type. The type \Rmnum{2} nature arises from entanglement with the algebra of the fluctuating causal complement region, that is, the commutant algebra
\begin{equation}    
       (\A^G)'=D(\{a,H_1-\hat{d}_1|a\in\A_{QFT}(\pi,0)\}'')
\end{equation}
which is also a type \Rmnum{2}${}_\infty$ factor. Here we have assumed the Haag duality \cite{Haag:1996hvx}: $\A_{QFT}(0,0)'=\A_{QFT}(\pi,0)$ in $\HH_{QFT}$.

While imposing a spectral lower bound on $\hat{d}_{1}$ would regulate the divergence in $\bra{0,0,0}_{\O}e^{-\hat{d}_1}\ket{0,0,0}_{\O}$ and yields a type \Rmnum{2}${}_1$ algebra as in CLPW, we refrain from doing so here because such a deformation is not applicable in the presence of multiple covariant observers, as discussed later.

\subsection{An action modeling the covariant observer}
In this subsection, we present a simple dynamical model that provides a concrete realization of the covariant observer which was introduced at the kinematical level, elucidating the physical origin of the fluctuating geodesic in terms of conserved charges and their quantization.

We begin by briefly reviewing the static observer considered in CLPW, which is described as a free particle carrying an internal clock degree of freedom, with action
\begin{equation}\label{eq:CLPW action}
    S=\int d\tau(\dot p q-(m+q))
\end{equation}
Here $\tau$ denotes the proper time, and $(p(\tau),q(\tau))$ is a conjugate pair describing the clock, kinematically independent of the spacetime position $x^\mu(\tau)$. Extremizing the action tells that the observer follows a timelike geodesic, with $\dot q=0$ and $\dot p=1$. The variable $p$, which coincides with the proper time $\tau$, is naturally interpreted as the charge associated with translations along the geodesic, while its conjugate $q$ acts as the corresponding Hamiltonian. Upon quantization, $[p,q]=i$ leads to the Hilbert space $L^2(\R)$, reflecting quantum uncertainty in the observer’s position along its worldline.

Motivated by this construction, we now seek a covariant generalization appropriate for an observer whose trajectory transforms nontrivially under the full de Sitter isometry group. Such a covariant observer should carry a set of conserved charges, each one corresponding to a generator of $SO(1,2)$, thereby encoding the choice of timelike geodesic in terms of integrals of motion rather than fixed background data. A natural choice of action is\footnote{We set $l_{dS}=1$; otherwise the charge associated with rotations should be rescaled by $l_{dS}$ to match dimensions.}
\begin{equation}\label{eq:action}
    S=\int d\tau (\dot p_A q^A-q^A\xi_A^\mu\dot x_\mu-m)
\end{equation}
where $(p^A,q^A)$ form conjugate pairs contracted with the Cartan–Killing metric of $SO(1,2)$, and $\xi_A^\mu$ denote the Killing vector fields generating $SO(1,2)$. The variables $q^A$ enter the gravitational constraints associated with $(\xi^A)^\mu$ as
\begin{equation}
\int_\Sigma\epsilon^\mu(\xi^A)^\nu T_{\mu\nu}=q^A
\end{equation}
in accordance with the gravitational constraints \eqref{eq: constraint single}.

Extremizing the action gives
\begin{equation}
    \dot q^A=0,\quad\dot p_A=\xi_A^\mu\dot x_\mu
\end{equation}
together with
\begin{equation}\label{eq:eom}
    m\dot x^\nu \nabla_\nu \dot x^\mu=q^AF_A^{\mu\nu}\dot x_\nu.
\end{equation}
In the regime $q^A\ll m$, which can be ensured by taking the mass sufficiently large, the right-hand side of \eqref{eq:eom} is parametrically suppressed. The observer therefore follows an approximately geodesic trajectory, with $q^A$ and $\dot p_A$ conserved. For the Killing generator corresponding to translations along the geodesic, one has $\dot p_A=-1$, whereas for the generators that only transform the geodesic without inducing translations, $\dot p_A=0$. 

Since the variables $p_A$ can be arbitrarily shifted without affecting the equations of motion, we may identify $(p_1,p_2,p_3)$ with the group parameters $(\phi,s,t)$ labeling the point $P_{(\phi,s,t)}$ on the geodesic $L_{(\phi,s)}$, and correspondingly rewrite $(q^1,q^2,q^3)$ as $(l,d_2,d_1)$. In this way, the conserved charges completely characterize the timelike geodesic, while the proper-time parameter $t$ specifies the observer’s location along it. The charges $q^A$ generate transformations associated with the corresponding de Sitter isometries.\footnote{Although the two actions discussed above are formally similar, their physical roles are distinct. The static observer introduces a single charge for translations along the worldline, whereas the covariant observer carries one charge for each Killing generator of $dS_2$. In the regime $q^A\ll m$, where the worldline coincides with the orbit of a boost, the covariant observer reduces to a direct generalization of the static one.}

Upon quantization, the canonical commutation relations give rise to the Hilbert space $L^2(SO(1,2))$ with the operators $\hat{\phi},\hat{s},\hat{t}$ and $\hat{l},\hat{d}_2,\hat{d}_1$ acting, furnishing the desired quantum reference frame.  Importantly, the fluctuating nature of the geodesic does not originate from quantizing small perturbations around a fixed classical worldline. Instead, it arises from the quantization of the conserved charges that label the geodesic itself. The observer is therefore described as a quantum superposition of classical geodesics related by de Sitter isometries, and the associated static patch fluctuates accordingly.

Finally, we note that the large-mass limit has been argued to be essential for obtaining a localized observer with a well-defined classical trajectory, suppressing wavepacket spreading\cite{Kolchmeyer:2024fly}. Here we adopt this regime to suppress the backreaction of the quantum reference frame on the observer’s trajectory, ensuring that the worldline remains approximately geodesic.

\subsection{Algebra for multiple covariant observers}
We now extend our framework to the presence of multiple covariant observers $\O_a,\O_b,\cdots$, to construct the algebra for one particular covariant observer, say $\O_a$. The fundamental principle remains: an observer $\O_a$ can only access degrees of freedom within its own causal patch $\M_a$. Since the QFT and  observers' degrees of freedom are related via gravitational constraints, we propose that $\O_a$ can access the following:
\begin{enumerate}
    \item The parameters $\phi_b,s_b$ specifying the location of $\O_b$'s geodesic, when it intersects $\M_a$.
    \item   The proper time interval $t_b$ along $\O_b$'s geodesic that lies within $\M_a$.
\end{enumerate}
Here we are holding the principle that from the perspective of $\O_a$, the other observers are viewed as part of the matter system, thus contribute to the algebra.

This leads to a decomposition of the Hilbert space $\HH_{\O_b}$. Based on the classical causal conditions \eqref{eq:geodesic-seen} and \eqref{eq:geodesic-seen-part}, we decompose $\HH_{\O_b}$ relative to the state $\ket{\phi_a,s_a,t_a}_{\O_a}$ of the observer $\O_a$:
\begin{equation}\label{eq:decompose hil}
    \HH_{\O_b}=\HH_{\O_b}^{(\phi_a,s_a)}\oplus \HH_{\U_b}^{(\phi_a,s_a)}
\end{equation}
where $\HH_{\O_b}^{(\phi_a,s_a)}$ is spanned by state $\ket{\phi_b,s_b,t_b}_{\O_b}$ with $|\phi_{-a+b}|<\frac{\pi}{2}$ and $t_b\in[F_{-a+b},G_{-a+b}]$ (i.e., the segment inside $\M_a$), and $\HH_{\U_b}^{(\phi_a,s_a)}$ is its orthogonal complement in $\HH_{\O_b}$. Physically, $\HH_{\O_b}^{(\phi_a,s_a)}$ contains all states that can be measured by $\O_a$. This clean decomposition is possible due to our use of an ideal QRF in which states with different $(\phi_b,s_b,t_b)$ are orthogonal.

Consequently, an observer $\O_a$ in state $\ket{\phi_a,s_a,t_a}_{\O_a}$ has access to the algebra of all bounded operators on the accessible sector
\begin{equation}\label{eq:other observer}
    \A_{\O_b}(\phi_a,s_a)=\B(\HH_{\O_b}^{(\phi_a,s_a)})\oplus \C\mathbf{1}_{\HH_{\U_b}^{(\phi_a,s_a)}}
\end{equation}
which is a type \Rmnum{1}${}_\infty$ algebra. Here $\B(\HH_{\O_b}^{(\phi_a,s_a)})$ denotes the algebra of all bounded operators on $\HH_{\O_b}^{(\phi_a,s_a)}$. The full kinematical algebra $\A_a$ for $\O_a$ is therefore generated by operators acting on its own Hilbert space $\hat{d}_{a,(\phi_a,s_a)}, \hat{t}_a$, the QFT algebra in its patch $\A_{QFT}(\phi,s)$, and the accessible algebras of all other observers:
\begin{equation}\label{eq:algebra multi}
 \A_{QFT}(\phi,s)\otimes\bigotimes_{b\neq a}\A_{\O_b}(\phi,s)\otimes\{\hat{d}_{a,(\phi,s)}, \hat{t}_a\}''\ket{\phi,s,t}_{\O_a}\bra{\phi,s,t}_{\O_a}.
\end{equation}
Here we adopt the notation that $\hat{d}_{a,i}$ and $\hat{l}_a$ act on the Hilbert space $\HH_{\O_a}$. One should note that the observer $\O_a$ itself and other observers contribute differently in \eqref{eq:algebra multi}, reflecting the observer-dependence of the resulting algebra. The center of the algebra
\begin{equation}
    \Z(\A_a)=\{\hat{\phi}_a,\hat{s}_a,\Theta(\frac{\pi}{2}-|\hat{\phi}_{-a+b}|)\Theta(\hat{t}_b-\hat{F}_{-a+b})\Theta(\hat{G}_{-a+b}-\hat{t}_b)|b\neq a\}''
\end{equation}
now includes projections that encode the causal relationships between the observers, indicating whether and when one observer is visible to another. Here $\Theta$ is the Heaviside step function, $\hat{F}_{-a+b}$ and $\hat{G}_{-a+b}$ are induced from the function $F_{-a+b},G_{-a+b}$ in \eqref{eq:geodesic-seen-part}.

The kinematical Hilbert space, comprising the degrees of freedom of both QFT and observers, is
\begin{equation}
    \HH_{kin}=\HH_{QFT}\otimes\bigotimes_{a}\HH_{\O_a}
\end{equation}
on which all kinematical operators act, with $a$ labeling different observers. The generators of the $SO(1,2)$ gauge symmetry are now summed over all observers: 
\begin{equation}
    \mathcal{C}_1=H_1+\sum_a\hat{d}_{a,1},\quad \mathcal{C}_2=H_2+\sum_a\hat{d}_{a,2},\quad \mathcal{D}=J+\sum_{a}\hat{l}_a.
\end{equation}
The isometry group is parameterized by
\begin{equation}\label{eq:full isometry}
      U(\phi,s,t)=e^{-i\phi \d}e^{-is \c_2}e^{-it\c_1}.
\end{equation}
Since the isometry preserves causal relations, $U(\phi,s,t)(\cdot)U^\dagger(\phi,s,t)$ generates an automorphism of $\A_a$. Introduce the dressing $D_a$ relative to the observer $\O_a$:
\begin{equation}
    D_a(a)=\int  d\phi dsdtU(\phi,s,t) aU^\dagger(\phi,s,t)\ket{\phi,s,t}_{\O_a}\bra{\phi,s ,t}_{\O_a}
\end{equation}
The gauge-invariant subalgebra $\A_a^G$ can be constructed similarly to the single observer case as\footnote{One can similarly define the dressing relative to any observer other than $\O_a$, the final gauge-invariant subalgebra $\A_a^G$, though, is unchanged.}
\begin{equation}\label{eq:algebra multi inv}
    \begin{aligned}
      \A^G_a
       =D_a(\A_{QFT}(0,0)\otimes\bigotimes_{b\neq a}\A_{\O_b}(0,0)\otimes\{\hat{d}_{a,1}\}'').\\
    \end{aligned}
\end{equation}
The physical interpretation is profound: the algebra $\A_a^G$ represents the observables accessible in a fluctuating region (due to $\O_a$'s quantized geodesic) where the number of internal degrees of freedom itself fluctuates (as other observers enter and exit the patch, in this way differs from ordinary field degrees of freedom).  
In this form, $\A_a^G$ is an observer-centered algebra. The observer $\O_a$ supplies the reference frame, while the QFT modes in $\M_a$ and the accessible parts of $\O_{b\neq a}$ are treated as the system. Choosing another observer $\O_b$ changes this split. Thus the family $\A_a^G$ is not merely a collection of different representations of a single fixed system algebra, but a family of observer-dependent descriptions in which the system/reference decomposition itself changes, giving a concrete realization of subsystem relativity \cite{AliAhmad:2021adn,Castro-Ruiz:2021vnq,delaHamette:2021oex,Hoehn:2023ehz,AliAhmad:2024wja}. 
The non-trivial center 
\begin{equation}
    \Z(\A^G_a)=\{\Theta(\frac{\pi}{2}-|\hat{\phi}_{-a+b}|)\Theta(\hat{t}_b-\hat{F}_{-a+b})\Theta(\hat{G}_{-a+b}-\hat{t}_b)|b\neq a\}''
\end{equation}
then tracks the dynamical causal relationships between observers in a quantum superposition. 

The algebra $\A_a^G$ is of type \Rmnum{2}${}_\infty$, as verified by the following construction. Define a state $\Psi_{\O_b}$ on $\A_{\O_b}(0,0)$ by
\begin{equation}
    \Psi_{\O_b}(a\otimes\alpha1_{\HH^{(0,0)}_{\U_b}})=tr_{\O_b}(e^{-\hat{d}_{b,1}}a)+\alpha,\quad a\in\B(\HH_{\O_b}^{(0,0)}),\alpha\in\C
\end{equation}
where  $tr_{\O_b}$ denotes the ordinary Hilbert space trace in $\HH_{\O_b}^{(0,0)}$. The state $\Psi_{\O_b}$ is thermal within $\HH_{\O_b}^{(0,0)}$ but vacuum-like within $\HH_{\U_b}^{(0,0)}$, with modular operator $\hat{d}'_{b,1}$ given by the restriction of $\hat{d}_{b,1}$ to $\HH_{\O_b}^{(0,0)}$. \footnote{Since the isometry preserves the causal relation, and the action of $\hat{d}_{a,1}$ preserves $\HH_{\O_b}^{(0,0)}$, it follows that $\HH_{\O_b}^{(0,0)}$ is also invariant under  $\hat{d}_{b,1}$. Thus the restriction $\hat{d}_{b,1}'$ is well-defined. The key distinction is that $\hat{d}'_{b,1}\in \B(\HH_{\O_b}^{(0,0)})$, whereas $\hat{d}_{b,1}$ is not confined to this subspace.\label{restriction generator}}  As $\A_{\O_b}(0,0)$ is defined on $\HH_{\O_b}^{(0,0)}$, the actions of $\hat{d}'_{b,1}$ and $\hat{d}_{b,1}$ on $\A_{\O_b}(0,0)$ coincide. Therefore, $\A_a^G$ is invariant under
\begin{equation}\label{restricted isometry}
      U'(\phi,s,t)=e^{-i\phi \d}e^{-is \c_2}e^{-it\c'_1}
\end{equation}
with
\begin{equation}
    \mathcal{C}'_1=H+\hat{d}_{a,1}+\sum_{b\neq a}\hat{d}'_{b,1}.
\end{equation}
Altogether, the state $\Psi_a$ on $\A_{QFT}(0,0)\otimes\bigotimes_{b\neq a}\A_{\O_b}(0,0)$, defined as
\begin{equation}\label{eq:Psi}
    \Psi_a=\Psi_{QFT}\otimes\bigotimes_{b\neq a} \Psi_{\O_b}
\end{equation}
satisfies the KMS condition with respect to the modular operator $H_1+\sum_{b\neq a}\hat{d}'_{b,1}$. Then a trace functional $Tr_a:\A_a^G\to\C$ can be defined as
\begin{equation}\label{trace}
    Tr_a(A)=\Psi_a(\bra{0,0,0}_{\O_a}e^{-\frac{\hat{d}_{a,1}}{2}}Ae^{-\frac{\hat{d}_{a,1}}{2}}\ket{0,0,0}_{\O_a})
\end{equation}
with $A\in\A_a^G$. Its normality, faithfulness and cyclic property (now use the invariance of $\A_a^G$ under $U'(\phi,s,t)$) can be verified similarly to the single observer case. Due to the divergence in the norm of $\Psi_{\O_b}$, we have $Tr(1)=\infty$, even if a spectral lower bound is imposed on $\hat{d}_{1}$.

Due to the nontrivial center, the trace on $\A^G_a$ is far from unique. Given the trace $Tr_a$, any central element defines a new trace via 
\begin{equation}
    Tr_a^z(A)=Tr_a(zA),\quad\forall z\in\Z(\A_a^G).
\end{equation}
For a normal state $\Psi$, if $\rho_\Psi$ is the density operator with respect to $Tr_a$, then the density operator for $Tr_a^z$ is $\rho z^{-1}$, resulting an entropy shift:
\begin{equation}
    S_{vN}^z(\Psi)=S_{vN}(\Psi)+\Psi(\log z). 
\end{equation}
The key to the construction of $Tr_a$ lies in the fact that the modular operator of $\Psi_{\O_b}$ acts the same as $\hat{d}_{b,1}$ on the algebra $\A_{\O_b}(0,0)$, which depends only on its behavior within $\HH_{\O_b}^{(0,0)}$. Another state with this property is of course
\begin{equation}
    \Psi'_{\O_b}(a)=tr_b(e^{-\hat{d}_{b,1}}a),\quad a\in\B(\HH_{\O_b})
\end{equation}
with $tr_b$ the ordinary Hilbert space trace in $\HH_{\O_b}$. The state $\Psi'_{\O_b}$ is thermal over the full Hilbert space $\HH_{\O_b}$ and has modular operator $\hat{d}_{b,1}$ on algebra $\A_{\O_b}(0,0)$. The KMS state $\Psi_a'$ with respect to  $H_1+\sum_{b\neq a}\hat{d}_{b,1}$ can be defined using $\Psi_{\O_b}'$, and the corresponding trace functional is\footnote{To avoid confusion, the trace $Tr_a$ is defined using the modified constraint $\c_1'$ and leads to the entropy $S_{vN}$; the trace $Tr_a'$ defined here uses the original constraint  $\c_1$ and leads to $S_{vN}'$. We primarily use $Tr_a$ and $S_{vN}$ in the main text because they yield an entropy that aligns with the generalized entropy.}
\begin{equation}\label{eq:another trace}
    Tr'_a(A)=\Psi'_a(\bra{0,0,0}_{\O_a}e^{-\frac{\hat{d}_{a,1}}{2}}Ae^{-\frac{\hat{d}_{a,1}}{2}}\ket{0,0,0}_{\O_a})
\end{equation}
with the entropy given by
\begin{equation}\label{eq:entropy difference}
    S_{vN}'(\Psi)=S_{vN}(\Psi)+\langle\sum_{b\neq a}(\hat{d}_{b,1}-\hat{d}'_{b,1})\rangle_\Psi.
\end{equation}

\subsection{Representation on a physical Hilbert space}
As mentioned in the introduction, the faithfulness of the representation of $\A_a^G$ remains an important question in the CLPW framework, suggesting the existence of an antipodal observer. This issue also arises in our construction. We first employ the perspective-neutral approach, constructing the gauge-invariant Hilbert space via group averaging \cite{Higuchi:1991tk,Higuchi:1991tm,Marolf:2008hg}, which supports a representation for $\A_a^G$. We then connect this construction to the Page-Wootters formalism \cite{Page:1983uc,Wootters:1984wfv} which describes physics from the perspective of a single observer. A key result is that the algebraic representation is faithful if and only if at least one other covariant observer is present.

The kinematical Hilbert space, comprising the degrees of freedom of both QFT and observers, is
\begin{equation}
    \HH_{kin}=\HH_{QFT}\otimes\bigotimes_{a}\HH_{\O_a}.
\end{equation}
We then define the physical Hilbert space $\HH_{phy}$ using the group-average inner product:
\begin{equation}
    \begin{aligned}
        (\psi|\psi')&=\int d\mu \bra{\psi}U(\phi,s,t)\ket{\psi'}\\
        &=\int d\phi dsdt\cosh s \bra{\psi}e^{-i\phi \d}e^{-is \c_2}e^{-it\c_1}\ket{\psi'}.
    \end{aligned}
\end{equation}
After mod out null states in this inner product, physical states $|\psi)$ are equivalence classes of kinematical states under the gauge group action
\begin{equation}
    |\psi)\sim U(\phi,s,t)|\psi)
\end{equation}
and satisfy the constraints $\c_1|\psi)=\c_2|\psi)=\d|\psi)=0$. Let $\zeta:\HH_{kin}\to\HH_{phy}$ denote the projection map, defined by $\zeta:\ket{\psi}\to|\psi)$. 

Any gauge-invariant operator $a\in\A_a^G$ is represented on $\HH_{phy}$ by $r(a)$, defined via
\begin{equation}
    r(a)\zeta\ket{\psi}=\zeta a\ket{\psi}
\end{equation}
The corresponding physical trace and density operator are defined by:
\begin{equation}
    Tr_a^{phy}=Tr_a\circ r^{-1},\quad (\phi|A|\phi)=Tr_a^{phy}(\rho A)
\end{equation}
leading to the von Neumann entropy:
 \begin{equation}
     S=-(\phi|\log \rho|\phi)=-Tr_a(r^{-1}(\rho)\log r^{-1}(\rho)).
 \end{equation}
 
The perspective-neutral framework can be related to a description from a specific observer's viewpoint via the Page-Wootters reduction. We define the reduction map $\r_{a}:\HH_{phy}\to\HH_{|a}$ to the perspective of observer $\O_a$ as 
\begin{equation}
    \ket{\psi_{|a}}=\mathcal{R}_a|\psi)=\bra{0,0,0}_{\O_a}\int d\mu U(\phi,s,t)\ket{\psi}.
\end{equation}
This map is unitary, and any physical state can be expressed as  \cite{DeVuyst:2024fxc,DeVuyst:2024khu}
\begin{equation}
    |\psi)=\zeta(\ket{\psi_{|a}}\otimes \ket{0,0,0}_{\O_a}).
\end{equation}
Crucially, if at least two observers $\O_a,\O_b$ are present, one can show the representation $r$ acts as: 
\begin{equation}
    r(a)=\r_b^\dagger (\pi_b a \r^\dagger_b)\r_b,\quad \forall a\in \A_a^G
\end{equation}
where we denote $\pi_b=\bra{0,0,0}_{\O_b}$. This can be verified as follows:
\begin{equation}
   \begin{aligned}
        \r_b r(a)|\psi)&=\r_b r(a)\zeta(\ket{\psi_{|b}}\otimes \ket{0,0,0}_{\O_b})=\r_b \zeta a(\ket{\psi_{|b}}\otimes \ket{0,0,0}_{\O_b})\\
        &=\bra{0,0,0}_{\O_b}\int d\mu U(\phi,s,t) a (\ket{\psi_{|b}}\otimes \ket{0,0,0}_{\O_b})\\
        &=\bra{0,0,0}_{\O_b}a\int d\mu U(\phi,s,t)  (\ket{\psi_{|b}}\otimes \ket{0,0,0}_{\O_b})\\
        &=\pi_b a |\psi)
   \end{aligned}
\end{equation}
Then $r(a)=0$ if and only if $\pi_b a=0$. Let $a=D_a(b)$, this means $\pi_b U(g)b U^\dagger(g)\ket{\psi}=0$ for all $g\in SO(1,2)$, and holds only if $a=0$. We thus indicates the representation $r$ is faithful.

Finally, as a consistency check, consider two observers with no QFT excitations. The constraints reduce to
\begin{equation}
    \hat{d}_{a,1}+\hat{d}_{b,1}=0,\quad\hat{d}_{a,2}+\hat{d}_{b,2}=0,\quad\hat{l}_{a}+\hat{l}_{b}=0
\end{equation}
forcing the observers to be antipodal, thus recovering the CLPW configuration. 

\section{Equality of algebraic and generalized entropy}\label{sec:entropy}
The type \Rmnum{2} nature of our gauge-invariant subalgebra guarantees a well-defined trace and, consequently, a notion of von Neumann entropy.  In this section, we calculate the entropy for semiclassical states \cite{Chandrasekaran:2022cip,Chandrasekaran:2022eqq,Kudler-Flam:2023qfl,Jensen:2023yxy,DeVuyst:2024fxc,DeVuyst:2024khu}, following the systematic procedure developed in \cite{DeVuyst:2024fxc,DeVuyst:2024khu} and the steps applied in \cite{Kirklin:2024gyl}.\footnote{Although the density operator beyond semiclassical states can also be obtained via the methods developed in the above references, it deviates from the main thread of the article, and it would be too complicate to do any specific computation. So we decide not to discuss it in this work.} By imposing a UV cutoff in QFT, and proposing a sensible quantum generalization of the first law for fluctuating region along with a fluctuating number of internal degrees of freedom, we demonstrate that the algebraic entropy coincides with the generalized entropy.

\subsection{ The von Neumann entropy for semiclassical states}
We now calculate the von Neumann entropy for semiclassical states. In this subsection, we write $a\in \A_{QFT}(0,0)\otimes\bigotimes_{b\neq a}\A_{\O_b}(0,0)$. The dressing $D_a$ relative to observer $\O_a$ is defined as
\begin{equation}
    D_a(a)=\int  d\phi dsdtU(\phi,s,t) aU^\dagger(\phi,s,t)\ket{\phi,s,t}_{\O_a}\bra{\phi,s ,t}_{\O_a}
\end{equation}
with $D_a(ae^{-i\hat{d}_{a,1}t})\in\A_a^G$.

A state is considered to be semiclassical if clock's state of an observer is sharply localized in time and essentially independent from the quantum fields and other observers, then serves as a reliable time reference for the rest of the system. Concretely, for a given state $\psi$ with density operator $\rho_\psi\in\A_a^G$, we require:
\begin{enumerate}
    \item   \textbf{Sharp time localization}: The clock itself must have small fluctuations, meaning its state is distinguishable at different times. Technically, this demands that the correlation function $Tr_a(\rho_\psi D( ae^{-i\hat{d}_{a,1}t})) $ is sharply peaked within  $|t|<\epsilon$.
    \item  \textbf{System-clock factorizability}: The clock must be approximately uncorrelated with the rest of the system (QFT and other observers). This ensures that the clock's reading can be used as a clean parameter without entangling with the system's state.  This is reflected in the factorization:\footnote{Here $\epsilon$ characterizes the clock time fluctuations, and $\approx$ indicates that we neglect terms of order higher than $\epsilon$. As $\epsilon$ is independent of $m$, the semiclassical states proposal is compatible with the $m\gg q^A$ regime which requires $\Delta t\gg \frac{1}{m}$.}
\begin{equation}
    Tr_a(\rho_\psi D_a( ae^{-i\hat{d}_{a,1}t}))\approx Tr_a(\rho_\psi D_a(a))Tr_a(\rho_\psi D_a(e^{-i\hat{d}_{a,1}t})),\quad \mathrm{if\;|t|<\epsilon}
\end{equation}
\end{enumerate}
Condition (1) ensures the clock is precise, while condition (2) ensures it is non-interfering. Together, the clock provides a approximately classical time parameter. This justifies the term semiclassical.

The density operator $\rho_\psi\in\A_a^G$ can be written as 
\begin{equation}
    \rho_\psi=\int_{-\infty}^\infty dt'D_a(e^{\frac{\hat{d}_{a,1}}{2}}e^{i\hat{d}_{a,1}t'}P(t')e^{\frac{\hat{d}_{a,1}}{2}})
\end{equation}
for some $P(t)\in \A_{QFT}(0)\otimes\bigotimes_{b\neq a}\A_{\O_b}(0,0)$. It obeys the relation
\begin{equation}\label{eq:density expect}
    \begin{aligned}
        Tr_a(\rho_\psi D_a(ae^{-i\hat{d}_{a,1}t}))&=\int_{-\infty}^\infty dt'\Psi_a(\bra{0,0,0}_{\O_a}e^{i\hat{d}_{a,1}t'}D_a(P(t'))e^{\frac{\hat{d}_{a,1}}{2}}D_a(a)e^{-\frac{\hat{d}_{a,1}}{2}}e^{-i\hat{d}_{a,1}t}\ket{0,0,0}_{\O_a})
        \\&=\Psi_a(P(t)e^{-(H_1+\sum_{b\neq a}\hat{d}'_{b,1})/2}ae^{(H_1+\sum_{b\neq a}\hat{d}'_{b,1})/2})
    \end{aligned}
\end{equation}
where the second equality uses the invariance of $D_a(P(t'))$ and $D_a(a)$ under $\c_1'$. We use $\c_1'$ rather than $\c_1$  because $\hat{d}_{b,1}\notin\A_{\O_b}(0,0)$, and thus its expectation value in $\Psi_a$ is undefined.

The first requirement of semiclassical state implies $P(t)$ is suppressed for $|t|>\epsilon$, while the second gives
\begin{equation}
    \begin{aligned}
       Tr_a(\rho_\psi D_a(ae^{-i\hat{d}_{a,1}t}))\approx \Psi_a(P(0)e^{-(H_1+\sum_{b\neq a}\hat{d}'_{b,1})/2}ae^{(H_1+\sum_{b\neq a}\hat{d}'_{b,1})/2})f(t)
    \end{aligned}
\end{equation}
where $f(t)=Tr_a(\rho_\psi D_a( e^{-i\hat{d}_{a,1}t}))$ is peaked in $|t|<\epsilon$. Then $P(t)\approx f(t)P(0)$ for $|t|<\epsilon$ due to the faithfulness of $\psi$, and therefore
\begin{equation}\label{eq:density halfroad}
    \rho_{\psi}=2\pi D_a( e^{\frac{\hat{d}_{a,1}}{2}}\tilde{f}(\hat{d}_{a,1})P(0)e^{\frac{\hat{d}_{a,1}}{2}})
\end{equation}
with
\begin{equation}
    \begin{aligned}
        \tilde{f}(\hat{d}_{a,1})&=\frac{1}{2\pi}\int dte^{i\hat{d}_{a,1}t}f(t)=\frac{1}{2\pi}\int dt e^{i\hat{d}_{a,1}t}Tr_a(\rho_\psi D_a(e^{-i\hat{d}_{a,1}t}))
    \end{aligned}
\end{equation}
The second requirement also leads to
\begin{equation}
    Tr_a(D_a(e^{i\hat{d}_{a,1}t})\rho_\psi D_a(e^{-i\hat{d}_{a,1}t} a))\approx Tr_a(\rho_\psi D_a(a))
\end{equation}
from which
\begin{equation}
    P(0)\approx e^{-i(H_1+\sum_{b\neq a}\hat{d}'_{b,1})t}P(0)e^{i(H_1+\sum_{b\neq a}\hat{d}'_{b,1})t},\quad |t|<\epsilon
\end{equation}
and in particular
\begin{equation}\label{eq:commute}
    [P(0),\tilde{f}(\hat{d}_{a,1})]\approx 0
\end{equation}
since $f(t)$ is peaked in $|t|<\epsilon$. To deal with the term $D_a(P(0))$ in \eqref{eq:density halfroad}, we observe that
\begin{equation}
    \begin{aligned}
       & Tr_a(\rho_\psi D_a(ab))
\\
     =&  \Psi_a(e^{(H_1+\sum_{b\neq a}\hat{d}'_{b,1})/2}be^{(-H_1+\sum_{b\neq a}\hat{d}'_{b,1})/2}P(0)e^{-(H_1+\sum_{b\neq a}\hat{d}'_{b,1})/2}ae^{(H_1+\sum_{b\neq a}\hat{d}'_{b,1})/2})\\
       =& \tilde{\Psi}_a\circ D_a(be^{-(H_1+\sum_{b\neq a}\hat{d}'_{b,1})/2}P(0)e^{-(H_1+\sum_{b\neq a}\hat{d}'_{b,1})/2}a)\\
        =&\tilde{\Psi}_a(D_a(b)D_a(e^{-(H_1+\sum_{b\neq a}\hat{d}'_{b,1})/2}P(0)e^{-(H_1+\sum_{b\neq a}\hat{d}'_{b,1})/2})D_a(a))
    \end{aligned}
\end{equation}
where the first equality is similar to \eqref{eq:density expect}, in the second equality we used the definition of $\Psi_a$ and introduced $\tilde{\Psi}_a$ defined by
\begin{equation}\label{eq:tilde psi}
    \tilde{\Psi}_a\circ D_a:=\Psi_{QFT}\otimes\bigotimes_{b\neq a}tr_{\O_b}. 
\end{equation}
From this, we identify
\begin{equation}\label{eq:relative modular}
    D_a(e^{-(H_1+\sum_{b\neq a}\hat{d}'_{b,1})/2}P(0)e^{-(H_1+\sum_{b\neq a}\hat{d}'_{b,1})/2})=\Delta_{\psi|\tilde{\Psi}_a}.
\end{equation}
where  $\Delta_{\psi|\tilde{\Psi}_a}$ is the relative modular operator of $D_a(\A_{QFT}(0)\otimes\bigotimes_{b\neq a}\A_{\O_{b}}(0,0))$ from state $\tilde{\Psi}_a$ to $\psi$.
Finally combining \eqref{eq:density halfroad}, \eqref{eq:commute} and \eqref{eq:relative modular}, and using the fact that $\c_1=\hat{d}_{a,1}+H_{1}+\sum_{b\neq a}\hat{d}'_{b,1}$ commutes with both $\tilde{f}(\hat{d}_{a,1})$ and $\Delta_{\psi|\tilde{\Psi}_a}\in\A_a^G$, we arrive at
\begin{equation}
  \begin{aligned}
        \rho_\psi\approx&2\pi D_a(e^{\frac{1}{2}(\hat{d}_{a,1}+H_{1}+\sum_{b\neq a}\hat{d}'_{b,1})}\tilde{f}(\hat{d}_{a,1}))\Delta_{\psi|\tilde{\Psi}_a}D_a(e^{\frac{1}{2}(\hat{d}_{a,1}+H_{1}+\sum_{b\neq a}\hat{d}'_{b,1})} )\\
        =&2\pi D_a(e^{\hat{d}_{a,1}+H_{1}+\sum_{b\neq a}\hat{d}'_{b,1}}\tilde{f}(\hat{d}_{a,1}))\Delta_{\psi|\tilde{\Psi}_a}.
  \end{aligned}
\end{equation}
This expression consists of three factors that all  commute approximately with each other.

With the above density matrix, the von Neumann entropy is then given by
\begin{equation}\label{eq:vN entropy}
    \begin{aligned}
        S_{vN}(\psi)\approx&-\log 2\pi-\langle D_a(\hat{d}_{a,1}+H_{1}+\sum_{b\neq a}\hat{d}'_{b,1})\rangle_\psi-\psi(\log\Delta_{\psi|\tilde{\Psi}_a})+S_{clock}(\psi)
    \end{aligned}
\end{equation}
where
$$S_{clock}(\psi)=\int_{-\infty}^\infty\bra{0,0,t_a}_{\O_a}D_a(\tilde{f}(\hat{d}_{a,1}))\log D_a( \tilde{f}(\hat{d}_{a,1}))\ket{0,0,t_a}_{\O_a}$$
represents the entropy contribution from the observer's clock fluctuations.

\subsection{Impose a UV cutoff}
To facilitate the evaluation of the entropy, we introduce some kind of UV regulator that renders the QFT algebra type \Rmnum{1}. This allows the relative modular operator to be decomposed in terms of density operators:
\begin{equation}
    \Delta_{\psi|\tilde{\Psi}_a}=\rho_{\tilde{\psi}}(\rho_{\tilde{\Psi}_a}')^{-1}
\end{equation}
Here, $\rho_{\tilde{\psi}}$ is the density operator associated with the state $\tilde{\psi}$, defined as the restriction of $\psi$ to the algebra $D_a(\A_{QFT}(0)\otimes\bigotimes_{b\neq a}\A_{\O_{b}}(0,0))$, and it generally differs from $\rho_\psi$. Meanwhile, $\rho'_{\tilde{\Psi}_a}$ is the density operator for the commutant algebra $\A_{QFT}(0,0)'$, which—assuming the Haag duality—coincides with $\A_{QFT}(\pi,0)$.
Using the definition of $\tilde{\Psi}_a$ in \eqref{eq:tilde psi} and the fact that $\Psi_{QFT}$ is thermal with respect to the one-sided QFT modular Hamiltonian $H_{\xi_1}'$, we have
\begin{equation}
    \rho'_{\tilde{\Psi}_a}=D_a(\rho_{\Psi_{QFT}}')=D_a(\frac{e^{-H_{\xi_1}'}}{Z})
\end{equation}
where the full boost generator decomposes as
\begin{equation}
    H_1=H_{\xi_1}-H_{\xi_1}'.
\end{equation}
Here, $H_{\xi_1}$ and $H_{\xi_1}'$ are the modular Hamiltonians associated with the boost Killing field $\xi_1$ in the $X^0-X^1$ plane, integrated over the Cauchy surfaces $\Sigma$ and $\Sigma'$ of the static patch $\M_0$ and its causal complement $\M_0'$, respectively:
\begin{equation}
    H_{\xi_1}=\int_\Sigma d\Sigma_a (\xi_1)_b T^{ab},\quad  H_{\xi_1}'=-\int_{\Sigma'} d\Sigma_a (\xi_1)_b T^{ab}.
\end{equation}

Substituting these into the entropy formula yields the simplified expression:
\begin{equation}
    S_{vN}(\psi)\approx -\langle D_a(\hat{d}_{a,1}+ H_{\xi_1}+\sum_{b\neq a}\hat{d}'_{b,1})\rangle_\psi +S_{clock}(\psi)-\psi(\log\rho_{\tilde{\psi}})+c\label{vonNeumann}
\end{equation}
where the state-independent constant $c$ is given by:
\begin{equation}
    c=-\log 2\pi-\log Z.
\end{equation}

\subsection{Quantum first law for the cosmological horizon}

The final step in establishing the equivalence between the algebraic and the generalized entropy is to relate the expectation value of the boost generator $\langle D_a(\hat{d}_{a,1}+ H_{\xi_1}+\sum_{b\neq a}\hat{d}'_{b,1})\rangle_\psi$ to the geometric perturbation of the cosmological horizon area. The standard first law applies to a fixed causal diamond. Our scenario, however, involves a \textit{fluctuating} static patch $\M_a$ whose locations are quantum variables and a \textit{fluctuating} number of internal degrees of freedom, which characterizes the entry and exit of other observers. We therefore propose a quantum generalization of the first law for such fluctuating subregions, which is an equality between the boost generator expectation and the quantum-average area perturbation $\frac{\langle A^{(2)}(a)\rangle_\psi}{4G_N}$.\footnote{While the notion of area is formally absent in $dS_2$, the following derivation is presented in a form that generalizes naturally to higher dimensions.}

Consider first a state $\psi_{(b_1,\cdots,b_n)}^{(\phi,s)}$ in which the observer $\O_a$ is localized on the fixed geodesic $L_{(\phi,s)}$, and only the observers $\O_{b_1},\cdots,\O_{b_n}$ intersect the static patch $\M_{(\phi,s)}$. For such a configuration with fixed degrees of freedom, the standard first law for a fixed subregion applies \cite{Gibbons:1977mu,Jensen:2023yxy}: 
\begin{equation}
     -\langle \hat{d}_{a,(\phi,s)}+H_{\xi_{(\phi,s)}}+\sum_{i}\hat{d}_{b_i,(\phi,s)}\rangle_{\psi_{(b_1,\cdots,b_n)}^{(\phi,s)}}=\frac{\langle A^{(2)}(\phi,s)\rangle_{\psi_{(b_1,\cdots,b_n)}^{(\phi,s)}}}{4G_N}.
\end{equation}
Here, $ A^{(2)}(\phi,s)$ denotes the second-order perturbation of the horizon area of $\M_{(\phi,s)}$, and $\xi_{(\phi,s)}$ is the Killing vector field that generates translation along $L_{(\phi,s)}$.

Recall that $\hat{d}'_{b,1}$ is defined as the restriction of $\hat{d}_{b,1}$ on $\HH_{\O_b}^{0,0}$, the Hilbert space that contains all states inside $\M_0$ (see the discussion in Footnote.~\ref{restriction generator}). Then $\hat{d}_{b,(\phi,s)}'$ can be defined similarly, and thus for the observers intersecting $\M_{(\phi,s)}$, we have $\langle\hat{d}'_{b_i,(\phi,s)}\rangle_{\psi_{(b_1,\cdots,b_n)}^{(\phi,s)}}=\langle\hat{d}_{b_i,(\phi,s)}\rangle_{\psi_{(b_1,\cdots,b_n)}^{(\phi,s)}}$, while for the observers $\O_c$ outside the patch, $\langle\hat{d}'_{c,(\phi,s)}\rangle_{\psi_{(b_1,\cdots,b_n)}^{(\phi,s)}}=0$ . The first law can therefore be rewritten as
\begin{equation}
     -\langle \hat{d}_{a,(\phi,s)}+H_{\xi_{(\phi,s)}}+\sum_{b\neq a}\hat{d}'_{b,(\phi,s)}\rangle_{\psi_{(b_1,\cdots,b_n)}^{(\phi,s)}}=\frac{\langle A^{(2)}(\phi,s)\rangle_{\psi_{(b_1,\cdots,b_n)}^{(\phi,s)}}}{4G_N}
\end{equation}
where the sum now runs over all other observers. This result can be generalized directly to the states $\psi^{(\phi,s)}$ in which $\O_a$ is still localized on $L_{(\phi,s)}$, but the geodesics of other observers are allowed to fluctuate. Such a state can be expanded as a superposition of states $\psi_{(b_1,\cdots,b_n)}^{(\phi,s)}$ with different sets $\{b_1,\cdots,b_n\}$, yielding a first law for a fixed subregion $\M_{(\phi,s)}$ but with a fluctuating number of internal degrees of freedom:
\begin{equation}\label{eq:first law half}
     -\langle \hat{d}_{a,(\phi,s)}+H_{\xi_{(\phi,s)}}+\sum_{b\neq a}\hat{d}'_{b,(\phi,s)}\rangle_{\psi^{(\phi,s)}}=\frac{\langle A^{(2)}(\phi,s)\rangle_{\psi^{(\phi,s)}}}{4G_N}.
\end{equation}

Now consider a fully general state $\psi$ in which the geodesic of $\O_a$ itself fluctuates. This state can be expanded as a superposition of states $\psi^{(\phi,s)}$ with different $(\phi,s)$. The quantum average of the left hand side of \eqref{eq:first law half} over the state of $\O_a$ naturally introduces the dressing $D$, since
\begin{equation}
     \langle D_a(\hat{d}_{a,1}+H_{\xi_1}+\sum_{b\neq a}\hat{d}'_{b,1})\rangle_{\psi^{(\phi,s)}}= \langle \hat{d}_{a,(\phi,s)}+H_{\xi_{(\phi,s)}}+\sum_{b\neq a}\hat{d}'_{b,(\phi,s)}\rangle_{\psi^{(\phi,s)}}.
\end{equation}
Moreover, it is natural to define $\langle A^{(2)}(a)\rangle_\psi$ as the quantum average of $\langle A^{(2)}(\phi,s)\rangle_{\psi^{(\phi,s)}}$ over the state of $\O_a$. We thus obtain the first law for a fully fluctuating region $\M_a$, with both its location and the number of its internal degrees of freedom subject to quantum fluctuations, 
\begin{equation}
    -\langle D_a(\hat{d}_{a,1}+H_{\xi_1}+\sum_{b\neq a}\hat{d}'_{b,1})\rangle_\psi=\frac{\langle A^{(2)}(a)\rangle_\psi}{4G_N}.\label{generalstate}
\end{equation}

Finally, with \eqref{generalstate} we establish the desired result that the von Neumann entropy \eqref{vonNeumann} equals the generalized entropy:
\begin{equation}\label{eq:vN=gen}
   \begin{aligned}
        S_{vN}(\psi)=&\frac{\langle A^{(2)}(a)\rangle_\psi}{4G_N}-\psi(\log\rho_{\tilde{\psi}})+S_{clock}(\psi)+c\\
       =&S_{gen}(\psi)+c.
   \end{aligned}
\end{equation}
Several remarks are in order:
\begin{enumerate}
    \item The semiclassical ansatz inherently distinguishes $\O_a$ from other degrees of freedom by assuming minimal entanglement, so that its contribution to the entropy  is dominated by $S_{clock}$.
    \item  Other covariant observers contribute to the von Neumann entropy in two ways, via their boost Hamiltonian, which affects the horizon area perturbation, and by introducing additional entanglement across the static patch. Note that $\rho_{\tilde{\psi}}$ is the density operator for $D(\A_{QFT}(0)\otimes\bigotimes_{b\neq a}\A_{\O_{b}}(0,0))$, so the term $-\psi(\log\rho_{\tilde{\psi}})$ captures the entanglement of both QFT modes and other covariant observers.  Collectively, these contributions preserve the generalized entropy formula. 
    \item The quantum nature of the observer's location is inherently incorporated: $\langle A^{(2)}(a)\rangle_\psi$ represents an average over all geodesic configurations. 
    \item Due to the nontrivial center of $\A_a^G$, the choice of trace functional is not unique, leading to different von Neumann entropies. For the alternative trace defined in \eqref{eq:another trace}, the associated entropy is
    \begin{equation}
        \begin{aligned}
            S_{vN}'(\psi)\approx& -\langle D_a(\hat{d}_{a,1}+ H_{\xi_1}+\sum_{b\neq a}\hat{d}_{b,1})\rangle_\psi +S_{clock}(\psi)-\psi(\log\rho_{\tilde{\psi}})+c\\
            =&\frac{\langle A^{(2)}(a)\rangle_\psi}{4G_N}-\psi(\log\rho_{\tilde{\psi}})+S_{clock}(\psi)+\langle\sum_{b\neq a}(\hat{d}_{b,1}-\hat{d}_{b,1}')\rangle_\psi+c .
        \end{aligned}
    \end{equation}
This can still be interpreted as a generalized entropy if we identify $\langle\sum_{b\neq a}(\hat{d}_{b,1}-\hat{d}_{b,1}')\rangle_\psi$ as part of the matter entropy contributed by other observers, which corresponds to the energy flux through the complementary patch $\M_0'$. This only reflects a shift of reference state in defining entropy. \footnote{A similar expression appears in \cite{Geng:2025bcb}, where the observer is constructed using a Goldstone vector field. In that context, the global KMS state cannot be obtained from a state that is thermal only within a finite patch by applying any finite number of local operations. This unitary inequivalence stems from the infinite number of field degrees of freedom and establishes the global KMS state as the more natural and physically complete construct. Therefore, it is natural to have such an energy term in the entropy there.} 
\item Gravitational entropy in our framework is inherently observer-dependent. This follows from the structure of the gauge-invariant algebra $\A_a^G$, in which the distinguished observer $\O_a$ contributes its own generator $\hat{d}_a$, whereas other observers are treated as dynamical matter components within the accessible algebra, see the discussion below \eqref{eq:algebra multi}. This intrinsic asymmetry provides a concrete realization of subsystem relativity  \cite{AliAhmad:2021adn,Castro-Ruiz:2021vnq,delaHamette:2021oex,Hoehn:2023ehz}. Moreover, the semiclassical prescription privileges the specific observer $\O_a$ by imposing severe restrictions on its clock state to recover the standard geometric entropy. 
\end{enumerate}

\section{Algebra in higher dimensions}\label{sec:higher dimension}
We now outline the extension of our framework to covariant observers in higher-dimensional de Sitter space $dS_d$ for $d\geq 3$. In these dimensions, a timelike geodesic is no longer uniquely specified by a single spacelike unit vector normal to a codimension-one hypersurface through the origin. Nevertheless, any timelike geodesic can be parameterized by the unique isometry that maps the reference geodesic $L_0=(\sinh \tau,\cosh \tau,0,\cdots,0)$ onto it.

We adopt a parametrization of $SO(1,d)$ analogous to the $SO(1,2)$ case: first, apply isometries that preserve $L_0$—namely, a boost in the $X^0-X^1$ plane and an $SO(d-1)$ rotation—then apply further boosts and rotations that change $L_0$ itself. Explicitly, we define
\begin{equation}
    g(\phi_i,s_i,\omega_j,t)=e^{-\sum_{i=2}^d\phi_i R_{i}}e^{-\sum_{i=2}^ds_i B_{i}}e^{-\sum_{2\leq i<j\leq d}\omega_{ij} R_{ij}}e^{-tB_1}
\end{equation}
where $B_i$ ($i=2,\cdots,d$) denotes the generator of boost in the $X^0-X^i$ plane, $R_i$ ($i=2,\cdots,d$) denotes the generator of rotation in the $X^1-X^i$ plane, $R_{ij}$ ($i,j=2,\cdots,d$) denotes generators of rotation in the $X^i-X^j$ plane, and $B_1$ denotes the generator of boost in the $X^0-X^1$ plane. With the composite parameters
\begin{equation}
    s=\sqrt{\sum_{i=2}^d s_i^2},\quad \phi=\sqrt{\sum_{i=2}^d \phi_i^2},
\end{equation}
 the left/right Haar measure in this parameterization are
\begin{equation}
    d\mu_L=d\mu_R=(\cosh s)^{d-1}(\cos\phi)^{d-2}\prod_{i=2}^dd\phi_i ds_i\prod_{2\leq i<j\leq d}d\omega_{ij}dt.
\end{equation}

In what follows, to simplify the notation, we will suppress the explicit summation over repeated indices $i,j$ and the product symbols in integration measures $d\phi_i ds_id\omega_{ij}dt$.  In the states as $\ket{\phi_i,s_i,\omega_{ij},t}$, the labels $ \phi_i,s_i,\omega_{ij},t$ collectively denote the entire set of parameters specifying the state, not its individual components; group elements $ U(\phi_i,s_i,\omega_{ij},t)$ and geodesics $L_{(\phi_i,s_i)}$ are denoted similarly.

Any timelike geodesic can then be labeled as $L_{(\phi_i,s_i)}$ ($i=2,\cdots,d$). The necessary and sufficient condition for a segment of this geodesic to lie within the static patch of $L_0$ is
\begin{equation}
    \cos\phi\geq0\quad\mathrm{i.e.}\quad\phi\in[-\frac{\pi}{2},\frac{\pi}{2}]
\end{equation}
with the accessible segment given by:
\begin{equation}
    \tanh\tau\in[-\frac{\cos\phi}{\cosh s-\sinh s\sin\phi},\frac{\cos\phi}{\cosh s+\sinh s\sin\phi}].
\end{equation}

Since $L_0$ is preserved by both the $X^0-X^1$ boost and the $SO(d-1)$ rotations, a complete reference frame requires not only a point along the geodesic but also a local orthonormal frame (a “pointer”) at that point. Quantum mechanically, this corresponds to equipping a covariant observer with both a clock and a full orthogonal frame. The observer’s Hilbert space is then
\begin{equation}
    \HH_\O\cong L^2(SO(1,d)),
\end{equation}
spanned by the vectors $\ket{\phi_i,s_i,\omega_{ij},t}_\O$ ($i,j=2,\cdots,d$ ), which describe an observer on the geodesic $L_{(\phi_i,s_i)}$ at proper time $t$, with its local frame oriented according to $\omega_{ij}$. We shall denote the generators to be $\hat{l}_i$, $\hat{d}_i$, $\hat{l}_{ij}$ and $\hat{d}_1$ accordingly, and the generators of the rotations and boost that preserve $L_{(\phi_i,s_i)}$ to be $\hat{l}_{ij,(\phi_i,s_i)}$ and $\hat{d}_{(\phi_i,s_i)}$ as in the case of $dS_2$. We choose a normalization that
\begin{equation}
\langle\phi_{i,1},s_{i,1},\omega_{ij,1},t_1|\phi_{i,2},s_{i,2},\omega_{ij,2},t_2\rangle_\O=\delta(\phi_{i,-1+2})\delta(s_{i,-1+2})\delta(\omega_{ij,-1+2})\delta(t_{-1+2})
\end{equation}
and
\begin{equation}
\mathbf{1}=\int d\phi_i ds_id\omega_{ij}dt\ket{\phi_i,s_i,\omega_{ij},t}_\O\bra{\phi_i,s_i,\omega_{ij},t}_\O.
\end{equation}
The group $SO(1,d)$ acts unitarily on $L^2(SO(1,d))$ via
\begin{equation}
   \begin{aligned}
        &U_{\O}(\phi_{i,1},s_{i,1},\omega_{ij,1},t_1)\ket{\phi_{i,2},s_{i,2},\omega_{ij,2},t_2}_{\O}\\
        =&\sqrt{\frac{(\cosh s_2)^{d-1}(\cos\phi_2)^{d-2}}{(\cosh s_{1+2})^{d-1}(\cos\phi_{1+2})^{d-2}}}\ket{\phi_{i,1+2},s_{i,1+2},\omega_{ij,1+2},t_{1+2}}_\O
   \end{aligned}
\end{equation}
where
\begin{equation}
    U_\O(\phi_i,s_i,\omega_{ij},t)=e^{-i\phi_i \hat{l}_{i}}e^{-is_i \hat{d}_{i}}e^{-i\omega_{ij} \hat{l}_{ij}}e^{-it\hat{d}_1}.
\end{equation}
We denote the unitary representation of the $SO(1,d)$ group in QFT Hilbert space as
\begin{equation}
    U_{QFT}(\phi_i,s_i,\omega_{ij},t)=e^{-i\phi_i J_{i}}e^{-is_i H_{i}}e^{-i\omega_{ij} J_{ij}}e^{-itH_1}
\end{equation}
Then the full $SO(1,2)$ constraints summed over all observers are
\begin{equation}
\mathcal{D}_i=J_i+\sum_{a}\hat{l}_{a,i},\quad\mathcal{C}_i=H_i+\sum_a\hat{d}_{a,i},\quad \d_{ij}=J_{ij}+\sum_a\hat{l}_{a,ij},\quad \c_1=H_1+\sum_a\hat{d}_{a,1}.
\end{equation}
The isometry group is parameterized by
\begin{equation}
    U(\phi_i,s_i,\omega_{ij},t)=e^{-i\phi_i \d_{i}}e^{-is_i \c_{i}}e^{-i\omega_{ij} \d_{ij}}e^{-it\c_1}.
\end{equation}

In the state $\ket{\phi_i,s_i,\omega_{ij},t}_{\O_a}$, we ascribe to the observer $\O_a$ the ability to access the following degrees of freedom:
\begin{enumerate}
    \item  The entire QFT algebra $\A_{QFT}(\phi_i,s_i)$ within its associated static patch $\M_{(\phi_i,s_i)}$.
    \item  Its own kinematic degrees of freedom, including the proper time  $\hat{t}_a$, the generator of translations along its geodesic,  $\hat{d}_{a,(\phi_i,s_i)}$, which enable evolution,  the orientation $\hat{\omega}_{a,ij}$ and the rotation generators $\hat{l}_{a,ij,(\phi_i,s_i)}$ of its orthogonal frame, which measure and rotate its own orthogonal frame.
    \item  The degrees of freedom of any other observer $\O_b$ whose geodesic intersects $\M_{(\phi_i,s_i)}$,  including the parameters $(\phi_{b,i},s_{b,i},\omega_{b,ij})$ and the proper time $t_b$ on the accessible segment. The associated algebra is $\A_{\O_b}(0)=\B(\HH^{0}_{\O_b})\otimes\C \mathbf{1}_{\HH_{\U_b}^0}$, where $\HH_{\O_b}^{0}$ is the subspace spanned by states of $\O_b$ inside $\M_0$ and $\HH_{\U_b}^0$ is its orthogonal complement in $\HH_{\O_b}$
\end{enumerate}
Therefore, the full $dS$-invariant algebra for one covariant observer $\O_a$ is an averaged version of the crossed product of $\A_{QFT}(0)\otimes\bigotimes_{b\neq a}\A_{\O_b}(0)$ over the group $\R\times SO(d-1)$:
\begin{equation}
    \A_a^G=D_a(\A_{QFT}(0)\otimes\bigotimes_{b\neq a}\A_{\O_b}(0)\otimes \{\hat{d}_{a,1}\}''\otimes\{\hat{l}_{a,ij}\}'')
\end{equation}
where we denote $\A_{QFT}(0)$ the algebra associated with $\M_{0}$, and $\A_{\O_b}(0)$ similarly.  The dressing isomorphism $D_a$, which represents the average over the group $SO(1,d)$, is defined as: 
\begin{equation}
    D_a(a)=\int  d\phi_i ds_id\omega_{ij}dt  U(\phi_i,s_i,\omega_{ij},t)a  U^\dagger(\phi_i,s_i,\omega_{ij},t)\ket{\phi_i,s_i,\omega_{ij},t}_{\O_a}\bra{\phi_i,s_i,\omega_{ij},t}_{\O_a}
\end{equation}
Following the results in \cite{Fewster:2024pur,AliAhmad:2024eun,Klinger:2026tws}, the algebra is type \Rmnum{2} as the group $\R\times SO(d-1)$ contains the modular automorphism group as a normal subgroup, and a trace functional can be constructed similarly:
\begin{equation}
    Tr_a(A)=\Psi_a(\bra{0_i,0_i,0_{ij},0}_{\O_a} e^{-\frac{\hat{d}_{a,1}}{2}}Ae^{-\frac{\hat{d}_{a,1}}{2}}\ket{0_i,0_i,0_{ij},0}_{\O_a}).
\end{equation}

Finally, the action for the covariant observer admits a straightforward generalization to higher dimensions,
\begin{equation}
    S=\int d\tau (\dot p_A q^A-q^A\xi_A^\mu\dot x_\mu-m)
\end{equation}
where $(p^A,q^A)$ now form conjugate pairs with indices contracted using the Cartan–Killing metric of $SO(1,d)$, and $\xi_A^\mu$ denote the Killing vector fields generating $SO(1,d)$. Interpretationally, it is worth pointing out that in \cite{Klinger:2026tws} the charges associated with the rotation subgroup $SO(d-1)$ admit a physical realization when the observer is modeled as an extended object of finite size rather than an ideal point particle.

\section{Conclusion and discussion\label{conclusion}}
In this work, we developed a framework that consistently incorporates the full set of second-order gravitational constraints associated with the de Sitter isometry group, while accommodating the presence of multiple observers along arbitrary timelike geodesics.  The central ingredient of our construction is the notion of \textit{covariant observer}, whose geodesic is treated not as a fixed background structure but as a dynamical object transforming covariantly under the isometry group. This perspective evades the symmetry breaking inherent in the original CLPW frameworks based on fixed reference geodesics.

Our main results can be summarized as follows:
\begin{enumerate}
    \item \textbf{The covariant observer as QRF}. In $dS_2$, the \textit{covariant observer}  carries a clock and moves along a dynamical geodesic, with degrees of freedom encoded in the group manifold $SO(1,2)$. Upon quantization, these form a Hilbert space $L^2(SO(1,2))$, providing a quantum reference frame for the full isometry group rather than a preferred subgroup $\R$. To clarify the physical origin of the fluctuating geodesic, we also introduced an explicit action model in which the observer carries conserved charges associated with each de Sitter isometry. In this picture, the quantum superposition of geodesics arises from the quantization of these charges, rather than from fluctuations around a fixed worldline. In higher dimensional $dS_d$, a complete QRF additionally requires an orthogonal frame, leading to the natural generalization $L^2(SO(1,d))$. 
    \item \textbf{An algebra of fluctuating region}. For a given covariant observer, we constructed an algebra of observables, which includes quantum field degrees of freedom within the observer’s static patch, but also the degrees of freedom of other observers whose worldlines intersect that patch, which are viewed as part of the matter system by the given observer.  Imposing the full de Sitter gauge constraints yields a $dS$-invariant subalgebra that can be understood as an averaged  modular crossed product. Importantly, this algebra is of type \Rmnum{2}, reflecting the fact that it should not be interpreted as the algebra of a single fixed region, but as the average of algebras over all possible static patches and geodesic configurations. 
    \item \textbf{Generalized entropy from algebraic entropy}. By imposing a UV cutoff and proposing a sensible quantum generalization of the first law suitable for fluctuating regions with a fluctuating number of internal degrees of freedom, we demonstrated that the von Neumann entropy of the $dS$-invariant type \Rmnum{2} algebra reproduces the generalized entropy for semiclassical states, including the contributions from covariant observers themselves. This provides further support for the view that gravitational entropy is intrinsically observer-dependent.
\end{enumerate}

Our framework suggests several promising directions for future research:
\begin{enumerate}
    \item \textbf{Beyond ideal QRFs}. A critical direction is to relax the assumption of an ideal QRF, as defined by \eqref{eq:ortho-complete} \cite{Hoehn:2019fsy,Hoehn:2020epv,DeVuyst:2024fxc,DeVuyst:2024khu}. Such an idealization is likely incompatible with the finiteness of de Sitter entropy \cite{Dyson:2002pf,Goheer:2002vf} and with nonperturbative gravitational effects, as discussed in ref.~\cite{Yang:2025lme}. This discussion parallels the restriction on the clock Hamiltonian spectrum imposed in the CLPW framework. Once the assumption of an ideal QRF is abandoned, the orthogonal decomposition \eqref{eq:decompose hil} underlying our construction no longer applies. Instead, we introduce the projection operator 
\begin{equation}
    \Pi_{\O_b}^{(\phi_a,s_a)}=\sum_{\substack{|\phi_{-a+b}|<\frac{\pi}{2}\\ t_b\in[F_{-a+b},G_{-a+b}]}}\ket{\phi_b,s_b,t_b}_{\O_b}\bra{\phi_b,s_b,t_b}_{\O_b}
\end{equation}    
which projects onto the portion of states contained in $\M_a$. A natural generalization of \eqref{eq:other observer} is then
    \begin{equation}
        \A_{\O_b}(\phi_a,s_a)=\Pi_{\O_b}^{(\phi_a,s_a)}\B(\HH_{\O_b})\Pi_{\O_b}^{(\phi_a,s_a)}
    \end{equation}
We expect that the analysis presented above extends straightforwardly to this generalized setting.

\item \textbf{Toward a top-down formulation}: It would be interesting to revisit the present multiple-observer construction from a more top-down perspective \cite{AliAhmad:2024wja}.  In this work, we have focused on the algebra accessible to a chosen observer which plays the role of the reference frame, while the QFT modes and other observers inside the patch are treated as the system. Choosing a different observer changes this split, giving a concrete realization of subsystem relativity \cite{AliAhmad:2021adn,Castro-Ruiz:2021vnq,delaHamette:2021oex,Hoehn:2023ehz}. This is closely related to the top-down crossed-product framework of \cite{AliAhmad:2024wja}, in which different crossed-product algebras may be viewed as local charts of a larger \(G\)-framed algebra.  Applying this perspective to covariant observers in de Sitter space may clarify which observables and entropies are genuinely frame-independent and which are intrinsically observer-dependent. A full implementation of this viewpoint is beyond the scope of this work and will be left open.

    \item   \textbf{Algebra for more general subregions in de Sitter space.}  In this work, we focused on the algebra associated with a fluctuating static patch defined by a covariant observer. It is natural to ask how this perspective extends to more general subregions in de Sitter space. 

A particularly instructive example is the causal region accessible to an observer with a finite lifetime, which is strictly smaller than the full static patch. The boost-like transformations preserving the horizon of such a finite-lifetime region act as modular flows under the geometric modular flow conjecture \cite{Jensen:2023yxy,Sorce:2023gio}, but they are not isometries of the de Sitter metric.   If one gauges only the full de Sitter isometry group, the algebra associated with a finite-lifetime region is therefore expected to remain of type \Rmnum{3}${}_1$. This provides an algebraic distinction between static patch and finite-lifetime patch, and may reflect the role of global gravitational effects associated with the future and past boundaries. Nevertheless, one may instead regard the emergence of type \Rmnum{2} algebras as a guiding principle, in which case additional symmetries beyond the de Sitter isometries would need to be incorporated. From this perspective, the corresponding transformation behavior of timelike trajectories must be taken into account in order to furnish a complete QRF.

Beyond single-observer regions, one may also consider multiple separated covariant observers whose static/finite-lifetime patches are distinct, tilted, and partially overlapping. In this setting, it is natural to associate algebras to intersections or unions of patches via algebraic intersection or union, allowing one to study constraints on entropy. In particular, for type \Rmnum{2} algebras, the strong subadditivity \cite{Luczak:2024nfi} can be applied to overlapping patches, potentially revealing aspects of the entanglement structure of de Sitter space, based on the ideas in \cite{Casini:2004bw,Abate:2024xyb}, which parallels the idea that gauging null translations leads to the generalized second law using the monotonicity of relative entropy \cite{Kirklin:2024gyl}.

More broadly, iterating the above construction may allow one to associate type \Rmnum{2} algebras—and thus well-defined algebraic entropies—to generic causally complete subregions in de Sitter space. This raises the possibility of a concrete algebraic realization of generalized entanglement wedges in de Sitter space \cite{penington2023entanglementwedgesgravitating,boussoTue2023,bousso2025fundamentalcomplementgravitating,Sahu:2025upe}, and provides a setting in which the proposed algebraic properties of such wedges \cite{Sahu:2025upe} could be tested directly.
    \item \textbf{Applications to general spacetimes}.  A key lesson of our construction is that the transformation properties of classical timelike trajectories provide a natural physical realization of quantum reference frames, instead of being tied to boundary degrees of freedom. An important direction is to explore how this framework extends beyond de Sitter space, to more general spatially closed spacetimes such as Schwarzschild–de Sitter black holes and cosmological backgrounds, as well as to asymptotically AdS or flat spacetimes. In particular, studying families of trajectories that cross black hole horizons may provide new insights into the connection between interior and exterior regions of black hole and the nature of the singularity.
    \item \textbf{Links to observer physics}. The role of observers in de Sitter space has been a central theme in numerous studies, see \cite{Kolchmeyer:2024fly,Silverstein:2022dfj,Maldacena:2024spf,Ivo:2025yek,Shi:2025amq,Jensen:2024dnl,Harlow:2025pvj,Abdalla:2025gzn,Akers:2025ahe,Tietto:2025oxn,Bousso:2025udh,Yang:2025lme,Guo:2025mwp,Blommaert:2025bgd,Chen:2025fwp,Li:2025fqz,Wei:2025guh,Engelhardt:2025azi,Antonini:2025ioh,Mertens:2025rpa,Irakleous:2025trr,Chen:2025jqm,Higginbotham:2025clp,Dulac:2025owj,Ali:2026vwk,Harlow:2026hky,Zhao:2026mpl,Nomura:2026igt,Klinger:2026kqj,Espindola:2026ekv,Espindola:2026uqa,Cui:2026bcd} for a selected list. Exploring connections between the action model \eqref{eq:action} introduced here and these approaches may help clarify the role of observers in quantum gravity.
    As a concrete example, following the spirit of \cite{Kolchmeyer:2024fly}, one may consider a fully quantized particle carrying a quantum reference frame QRF $L^2(SO(1,2))$ in order to study effects beyond the strict large-mass limit. In such a framework, the canonical variables $(q^A,p^A)$ are expected to continue to encode effective trajectories and thus define operational notions of observability. This setting allows one to probe the recoil of observer trajectories via out-of-time-order correlators, where the presence of a horizon suggests saturation of the chaos bound. More generally, a path-integral quantization of the action \eqref{eq:action} could offer a starting point for investigating nonperturbative gravitational effects, along the lines explored in related contexts \cite{Yang:2025lme,Maldacena:2024spf,Ivo:2025yek,Shi:2025amq,Chen:2025jqm}.  

    Moreover, our construction may have implications for holography in de Sitter space   \cite{Susskind:2021omt,Susskind:2021dfc,Shaghoulian:2022fop,Kawamoto:2023nki,Hao:2024nhd,Ruan:2025uhl}. In particular, the emergence of de Sitter–invariant structures through averaging over the isometry group suggests a possible organizing principle for static-patch holography, in which observer dependence and symmetry averaging play a central role.
\end{enumerate}

Finally, let us comment on the relation between our work  with several existing approaches: 
\begin{enumerate}
    \item Multiple observers have been considered in the framework of CLPW in Ref.~\cite{DeVuyst:2024fxc,DeVuyst:2024khu}, where the observers move along fixed orbits of the boost symmetry, not necessarily geodesics. While considering a general trajectories is valuable, this approach still breaks the symmetry down to $\R\times SO(d-1)$. When multiple observers are present, these works assume each observer can access some clock degrees of freedom of others as an operational specification. As the trajectories are either entirely inside or outside a given static patch, the possible application of timelike tube theorem becomes essentially  trivial.
    \item Gauging full $dS$ isometry has been considered in Ref.~\cite{Kaplan:2024xyk}, which used a field rather than an observer as a reference frame, thus avoiding symmetry breaking. However, the gauge-invariant algebra in that framework is inherently smeared and does not take the crossed product form, precluding a discussion of von Neumann entropy.
    \item The dynamical observer model introduced by Kolchmeyer and Liu (KL) in Ref.~\cite{Kolchmeyer:2024fly} shares several conceptual similarities with our approach. In that framework, based on the action \eqref{eq:CLPW action}, an observer is modeled as a fully quantized relativistic particle with clock degrees of freedom. Since the particle wavefunction is spread over the entire de Sitter space, local QFT operators smeared over a Cauchy slice can be dressed to the observer to form gauge-invariant operators. The resulting algebra thus takes the form of a direct integral of \Rmnum{1}${}_\infty$ factors. However, as the dynamical observer is intrinsically delocalized, the framework does not provide a natural criterion for identifying which observables are operationally accessible to a given observer. In particular, it does not single out an cosmological horizon, and the entropy of the resulting type \Rmnum{1} algebra is not naturally associated with the generalized entropy of a horizon. In contrast, based on the action \eqref{eq:action}, our construction preserves the notion of an observer-dependent horizon while allowing the observer’s geodesic to fluctuate covariantly, leading to a de Sitter–invariant type \Rmnum{2} algebra and an entropy that admits an interpretation as generalized entropy, in closer spirit to CLPW. 
\end{enumerate}

\acknowledgments
 We would like to thank Shan-Ming Ruan for inspiring discussions and thank Hao Geng, Philipp A. H\"ohn, Zezhou Hu, Xin-cheng Mao, Tomonori Ugajin, Jun-kai Wang, Zixia Wei, Yu-ting Wen, Zhenbin Yang and Zhi-jun Yin for the valuable suggestions on the manuscript. This research is supported in part by NSFC Grant No. 12275004, 12588101.
\appendix

\bibliographystyle{JHEP}
\bibliography{biblio.bib}
\end{document}